\documentclass[10pt, showpacs,amsmath,amssymb,aps,twocolumn,superscriptaddress,prx, footinbib]{revtex4-2} 

\usepackage{graphicx}
\newcommand{\paperfigureheight}{0.33005\textwidth}
\usepackage{xcolor}

\usepackage{mathtools}
\usepackage{amsthm}
\usepackage{grffile} 

\usepackage{ulem} 
\usepackage{comment}

\usepackage{dcolumn}
\usepackage{bm}
\usepackage[colorlinks=true]{hyperref}

\usepackage{cleveref}
\crefname{appendix}{App.}{Apps.}
\crefname{equation}{Eq.}{Eqs.}
\crefname{figure}{Fig.}{Figs.}
\crefname{table}{Tab.}{Tabs.}
\crefname{section}{Sec.}{Secs.}

\hypersetup{
    colorlinks=true,
    linkcolor=blue,    
    citecolor=blue,   
    urlcolor=blue, 
}

\usepackage{color}
\definecolor{ForestGreen}{RGB}{34, 139, 34}

\definecolor{CrimsonRed}{RGB}{220, 20, 60}

\newcommand{\affA}{Department of Physics, Princeton University, Princeton, New Jersey 08544, USA}
\newcommand{\affB}{Department of Theoretical Physics, University of Geneva, Geneva, Switzerland}

\newcommand{\affC}{Institute of Physics, \'Ecole Polytechnique F\'ed\'erale de Lausanne (EPFL), 1015 Lausanne, Switzerland}

\newcommand{\affD}{Google Research, Brandschenkestrasse 150, 8002 Z\"urich, Switzerland}
\newcommand{\affE}{Max-Planck-Institut f\"ur Quantenoptik, Hans-Kopfermann-Straße 1, D-85748 Garching, Germany}
\newcommand{\affF}{Munich Center for Quantum Science and Technology (MCQST), Schellingstraße 4, D-80799 Munich, Germany}

\begin{document}

\title{Single-shot ground-state preparation via strongly coupled baths}

\author{Liang-Hong Mo}
\affiliation{\affA}

\author{Jerome Lloyd}
\affiliation{\affB}
\affiliation{\affE}
\affiliation{\affF}

\author{Yuxuan Zhang}
\affiliation{\affA}
\affiliation{\affC}

\author{Dmitry A. Abanin}
 \affiliation{\affA}
\affiliation{\affC}
\affiliation{\affD}
 	\author{Biao Lian}
 \affiliation{\affA}

\begin{abstract}

Preparing ground and thermal states of quantum many-body systems is a central task in quantum simulation. Dissipative protocols have emerged as a promising solution for both near-term and fault-tolerant hardware, but most constructions rely on weak system-bath coupling, leading to slow cooling. Here, we construct cooling protocols at strong system-bath coupling, and demonstrate accurate state preparation within even a single cooling cycle. Our protocols combine dynamically ramped ancilla energy levels with time-dependent couplings to optimize the trade-off between cooling error and physical time. For a general class of free-fermion systems, we reduce the dynamics exactly to a two-level scattering problem: ground-state cooling becomes a control problem of maximizing resonant excitation transfer while suppressing non-adiabatic heating. We then construct a family of ``M\"obius--Ingham'' protocols, which reach optimal error scaling for gapless spectra and almost-exponential suppression of the leading heating error for gapped spectra, as a function of the cycle period. In our free-fermion benchmarks, M\"obius--Ingham cooling lies on the Pareto frontier between steady-state error and cooling time among the controls tested. We further demonstrate its effectiveness in interacting Ising and $t$--$V$ chains and the two-dimensional Fermi--Hubbard model. Finally, we extend our construction to Gibbs state preparation, showing how free-fermion thermal states can be prepared in a single cycle.
\end{abstract}
\maketitle

\begin{figure*}[t]
\centering
\includegraphics[height=\paperfigureheight]{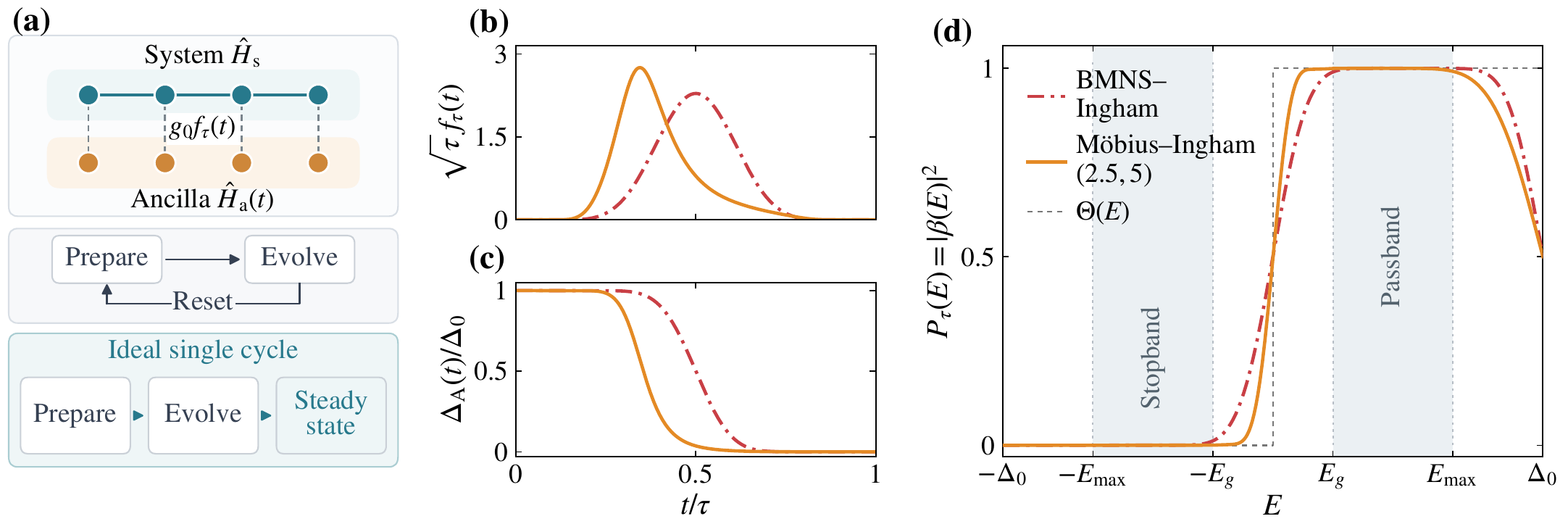}
\caption{Illustration of cooling cycle and control functions.
(a) System-ancilla coupling, the prepare-evolve-reset cycle, and the ideal single-cycle limit. Panels (b-d) are calculated for the BMNS--Ingham control and M\"obius--Ingham control with $(c_1,c_2)=(2.5,5)$ (see legend in (d)). (b) The corresponding normalized coupling function $\sqrt{\tau}f_\tau(t)$ and (c) ancilla ramping $\Delta_{\mathrm A}(t)/\Delta_0$ versus $t/\tau$ satisfying the optimal matching condition in Eq.~\eqref{eq:optimal_matching_condition}. (d) Transfer probability $P_\tau(E)$ compared with the ideal step $\Theta(E)$ (gray dashed line). The shaded intervals mark the stopband (left) and the passband (right). The parameters are $\tau=60$, $E_g=0.5$, $E_{\max}=1.5$, $\Delta_0=E_{\max}/(1-\alpha),\alpha=\frac{1}{3}$, $\Gamma_0=25$, and $|V|^2=1/2$.  }
\label{fig:cooling_protocol}
\end{figure*}

\section{Introduction}
\label{sec:introduction}
Preparing ground and thermal states of a quantum many-body system is a central task in quantum simulation and quantum computation~\cite{Bloch_2008,cirac2012goals, Quantumsimulation2014,cerezo2021variational}. Engineered dissipative dynamics offers a route toward this goal~\cite{PZpreparestateqmp2008,verstraete2009quantum}.
 In ancilla-assisted cooling, energy and entropy are transferred to auxiliary degrees of freedom that are reset between cycles~\cite{PhysRevResearch.2.023214,Puenteancillaresetting,DiluteMeasurementInducedCooling2024,2024Sci...383.1332M,2024Quant...8.1505M,motlagh2024groundstatepreparationdynamical,GaugedCooling2025,Marti_2025coolingfermion,seki2026dissipative,langbehn2026universalcoolingquantumsystems,2026modelagnosticcoolingalgorithmsstrongly}. A central question is how to reach the target state with high accuracy while keeping the total preparation time short.

Much of the theoretical understanding of this question has been developed in the perturbative weak system--bath coupling regime, where the reduced dynamics can be described by effective Lindblad equations or kinetic equations~\cite{PRXQuantum.6.010361,simplecasestudycool2025,hagan2025thermodynamiccostignorancethermal}. Lindblad generators can also be constructed directly to prepare thermal and ground states~\cite{DingPhysRevResearch.6.033147,chen2025efficient}, with rigorous convergence guarantees established for specific classes of systems~\cite{YTzhanPRX2026,bergamaschi2026fastmixingquantumspin,Ding2026,Hahn_2026}.

In the weak-coupling regime, cooling and heating are set by Fermi's golden-rule-level scattering rates 
which can be engineered by controlling the ancilla frequencies and system-bath coupling profile. This principle is behind the design of Gaussian filters for Gibbs-state preparation~\cite{chen2025efficient,Jeromegaussian2026}, and time-dependent coupling protocols for ground state preparation~\cite{PRXQuantum.6.010361, DingPhysRevResearch.6.033147}. These protocols perform well in the weak-coupling regime, but their leading cooling rates are quadratic in the coupling strength, leading to long cooling timescales inversely proportional to the squared coupling.

Strong coupling may circumvent this perturbative timescale and has recently been explored in Refs.~\cite{PhysRevA.105.052601,wang2026lindbladapproximationrigorousguarantees}. However, in this regime, the leading-order expansion is insufficient, the dynamics becomes intrinsically non-Markovian, and the contributions from all orders must be accounted for. This makes it substantially more challenging to characterize the cooling dynamics and optimize the control functions.  This raises a natural question:\textit{What principles govern the design of efficient ground- and thermal-state preparation protocols in the strong-coupling regime?}

To tackle this difficult problem, we first focus on a class of free-fermion systems strongly coupled to dynamical ancillas.  We consider protocols in which both the system--ancilla coupling and the ancilla energy levels vary in time. The full dynamics reduces exactly to independent two-level scattering problems, providing a non-perturbative framework for control design ---closely related to the inverse scattering problem in nuclear magnetic resonance pulse design \cite{epstein2004introduction}. For ground-state preparation, the optimization reduces to designing control functions that produce a step-like transfer probability over the system spectrum (see Figure \ref{fig:cooling_protocol}d). Motivated by Landau-Zener dynamics~\cite{zener1932non}, we construct cooling protocols by ramping the ancilla energies through resonance with the system modes to transfer excitations from the system to the ancillas. Our control design aims to maximize resonant excitation transfer while suppressing nonadiabatic heating. We further extend this framework to finite-temperature Gibbs-state preparation by balancing heating and cooling.

Our main results are as follows:
\begin{enumerate}
    \item \textit{Exact dynamics and optimal matching.}  Within the exact two-level reduction, we show how smooth coupling functions and monotonic ancilla ramps suppress the ground-state cooling error at strong coupling and large cycle periods $\tau$, producing a transfer probability that approximates the ideal step function. Motivated by the Landau--Zener approximation, we introduce an optimal matching condition between the coupling function and the ancilla ramp to optimize resonant excitation transfer across the system band. This condition determines the ramp from the coupling function, leaving a single control function to be optimized.
\item \textit{Control functions and error scaling.} For general systems whose low-energy spectra are not known in advance, we further optimize the coupling function to suppress nonadiabatic heating and construct a family of control functions, which we call M\"obius--Ingham functions. This family of functions achieves \textit{almost-exponential} suppression of the leading gapped heating error and \textit{optimal} gapless error scaling simultaneously.
\item \textit{Benchmarks beyond the analytical setting.}  We benchmark our ground-state cooling protocols in two-dimensional free-fermion systems, interacting Ising and $t$--$V$ chains, and a two-dimensional Fermi--Hubbard model. Across different models and different spectra, we find that the ground-state fidelity generically approaches its steady-state value within one or two cycles.

\item \textit{Gibbs-state preparation beyond weak coupling}. Finally, we further extend this framework to finite-temperature Gibbs-state preparation by balancing heating and cooling. We identify a detailed balance condition that can be approximately satisfied by the Demkov--Kunike (DK) protocol, with the target temperature set by the width of the coupling profile.
\end{enumerate}

\section{Cooling setup and optimization objective}

\begin{figure*}[t]
\centering
\includegraphics[height=\paperfigureheight]{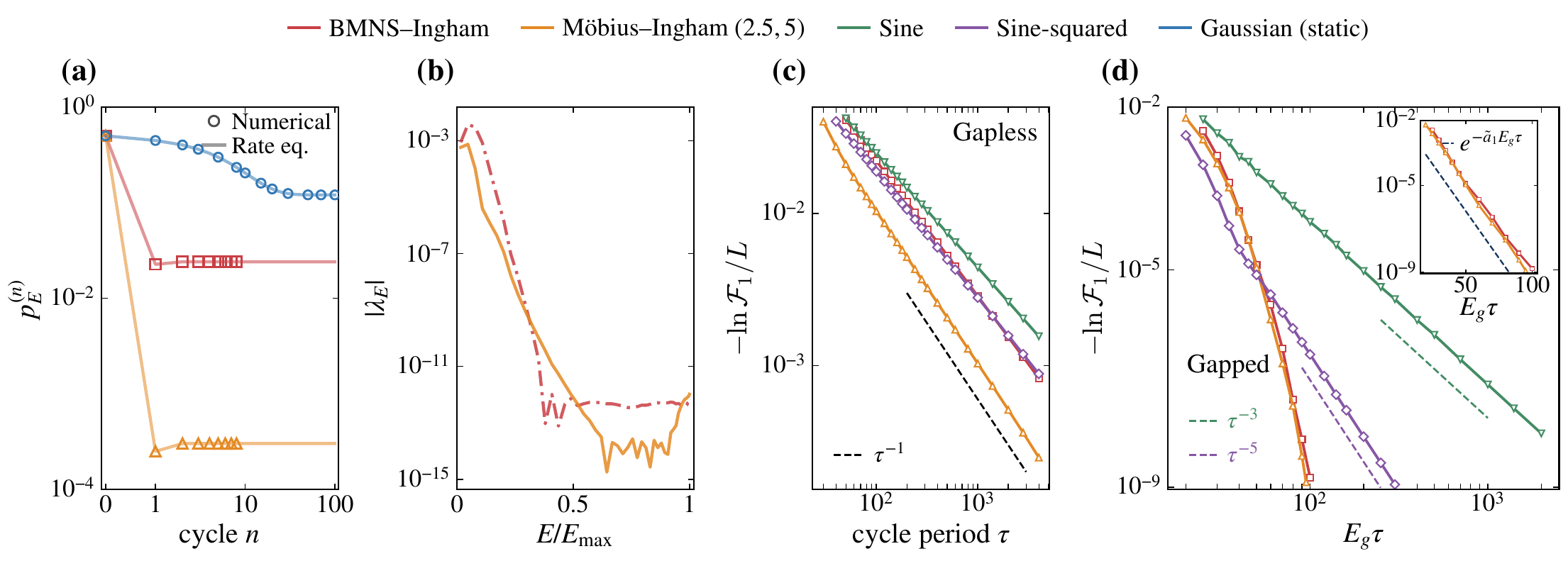}

\caption{ Single-cycle cooling dynamics and error scaling
in the SSH model.
(a) Pair excitation density $p_E^{(n)}$ versus cycle number $n$ for the chiral pair around the Fermi surface in the gapless chain. The BMNS--Ingham and M\"obius--Ingham with $(c_1,c_2)=(2.5,5)$ are calculated at $\tau=200$. The Gaussian protocol~\cite{Jeromegaussian2026} has static ancilla energies $\pm\Delta_{\rm G}$, $\Delta_{\rm G}\simeq1$, $\beta=10$, $g_0=1$,
and $\tau_{\rm G}\simeq15.8$. Markers denote the full single-particle evolution and solid lines are from the two-level rate equation. Here we use $L_a=L=100$. (b)  Single-cycle condition error $|\lambda_E|$ versus energy for the same chain at $\tau=200$. (c,d) Single-cycle log-fidelity density $-\ln\mathcal F_1/L$ for chains with $L=5000$
system sites and $L_a=L$ ancilla modes in (c) gapless and (d) gapped chains. Dashed lines indicate $\tau^{-1}$ in (c) and $\tau^{-3}$ and $\tau^{-5}$ in (d). The inset in (d) shows the two Ingham protocols on a semilogarithmic scale, with
$\exp(-\tilde a_1E_g\tau)$ as a guide to the eye. The gapless and gapped chains have $(t_1^{(s)},t_2^{(s)})=(1,1)$ and $(1,0.5)$, respectively. All chains have open boundaries and initial covariance $C_0=I/2$. The dynamical controls have the same $\Gamma_0$, $\alpha$, and $|V|^2$ as in Fig.~\ref{fig:cooling_protocol}. }
\label{fig:filter_error_scaling}

\end{figure*}

We now introduce our dissipative cooling protocol, which follows the setup of Ref.~\cite{PRXQuantum.6.010361} . We consider a system with a general many-body Hamiltonian $\hat H_{\mathrm s}$ coupled to resettable ancillas, as illustrated in Fig.~\ref{fig:cooling_protocol}(a). Our goal is to design the joint dynamics so that the system approaches a steady state close to the target state within a short physical time.

Specifically, each evolution cycle has period $\tau$ and consists of three stages. First, the ancillas are prepared in the ground state (density matrix) $\hat\rho_a(0)$ of $\hat H_a(0)$. Second, the system and ancillas evolve jointly under $\hat H(t) =\hat H_{\mathrm s}+\hat H_{\mathrm a}(t)+\hat H_{\mathrm c}(t)$. Finally, the ancillas are measured and reset before the next cycle, discarding the measurement outcomes. The resulting evolution of the system's density matrix $\hat \rho_s$ in the $(m+1)$-th cycle is described by the quantum channel evolution
\begin{align}\label{Eq:channel}
     \hat \rho_s^{(m+1)}= \mathcal{E}(\hat \rho_s^{(m)})= \mathrm{tr}_a[\hat{ \mathcal{   U}}^{\tau}(\hat \rho_s^{(m)}\otimes \hat \rho_a(0))  (\hat{ \mathcal{   U}}^{\tau})^{\dagger}],
\end{align}
where the one-cycle unitary is  $\hat{ \mathcal{U}}^{\tau} = \mathcal{T}\exp\big[-i \int_0^{\tau}dt\ \hat H(t)\big]$ with system-ancilla coupling
\begin{align}
    \hat H_c(t) = g_0  f_{\tau}(t) \sum_j\bigg ( \hat S_j\hat A_j^{\dagger }+\mathrm{h.c.}\bigg),
\end{align}
where $\hat S_j$ and $\hat A_j$ act on the system and ancilla, respectively, and the coupling function $f_\tau(t)$ is nonzero only for $t\in[0,\tau]$. We assume $f_\tau(t)$ is real (see the justification in \cref{subsubsec:LZ}) and $L_2$ normalized\footnote{
For a two-level transition, a time-dependent coupling phase can be removed into the ancilla ramping by a rotating-frame transformation.
}, $\int_0^\tau |f_\tau(t)|^2\,dt=1$, so that $g_0^2$ determines the coupling strength. We denote its Fourier transform as $\widetilde f_\tau(\omega)=\int_0^\tau f_\tau(t)e^{i\omega t}dt$.

We consider protocols in which the ancilla energy levels vary during each cooling cycle. For free-fermion ancillas, we adopt
\begin{equation}
\hat H_{\mathrm a}(t)
=\Delta_{\mathrm A}(t)\sum_i
\left(\hat s_i^\dagger\hat s_i-\hat r_i^\dagger\hat r_i\right),\qquad \Delta_{\mathrm A}(t)\ge 0\ ,
\label{eq:dynamical_ancilla_hamiltonian}
\end{equation}
where $i$ labels a pair of sink and source ancilla fermion modes, with energies $+\Delta_{\mathrm A}(t)$ and $-\Delta_{\mathrm A}(t)$, and annihilation operators $\hat s_i$ and $\hat r_i$, respectively. In general, the number of ancillas $L_a= zL$ is of the same order as the number of system fermion modes (or qubits) $L$. For all systems considered in this paper, $z\le 2$. At the beginning of each cycle, the sink is empty and the source is occupied. In our dynamical protocols, the ramping function $\Delta_{\mathrm A}(t)$ decreases monotonically from $\Delta_0>0$ to zero over the cycle period $\tau$.  For this Hamiltonian, a rotating-frame transformation can shift this time dependence into the phases of the corresponding system--ancilla couplings, giving an equivalent description with fixed ancilla energies~\cite{DingPhysRevResearch.6.033147,Ong_2026}.
Spin ancillas can be defined similarly using a time-dependent level splitting, e.g., $\hat H_{\mathrm a}(t)=-\Delta_{\mathrm A}(t)/2\sum_i\sigma^z_{i}$.

The ancilla ramping is motivated by Landau--Zener dynamics~\cite{zener1932non}. A transition between system many-body eigenstates $m$ and $n$ with energies $E_m>E_n$ becomes resonant with an ancilla excitation when $\Delta_{\mathrm A}(t)=E_m-E_n$. The coupling can then transfer the system excitation energy to the ancilla, which is subsequently removed by resetting. 
Efficient cooling requires enhancing this resonant transfer while suppressing nonadiabatic heating during the evolution. Our control problem is therefore to choose the coupling function $f_\tau(t)$ and ramping function $\Delta_{\mathrm A}(t)$ to achieve both goals.

We first focus on ground-state preparation and quantify the cooling accuracy after $m$ cycles by the ground-state fidelity
\begin{equation}
\mathcal F_m
=
\langle\mathrm{GS}|
\hat\rho_{\mathrm s}^{(m)}
|\mathrm{GS}\rangle\ ,
\end{equation}
where $|\mathrm{GS}\rangle$ is the system's ground state. It is worth noting that many-body fidelity provides a stringent test of ground-state preparation, but it can be evaluated directly for the free-fermion systems considered below.  In practice, reproducing local properties
of the ground state may already be sufficient.

At fixed coupling strength, the control problem is to reach a given fidelity in the shortest possible time. The best attainable trade-off between fidelity and evolution time defines the \textit{Pareto frontier}~\cite{miettinen1999nonlinear}. In particular, we aim to design protocols that achieve a small ground-state error comparable to the steady-state error after just one cooling cycle.

\section{Cooling dynamics and control design in Free-Fermion Systems}\label{sec:cooling-mechanism}

We now analyze how the coupling function and ancilla ramp can be designed for rapid ground-state preparation in free-fermion systems. We consider a class of models whose cooling dynamics can be exactly decomposed into independent two-level systems. This allows us to analyze cooling dynamics without assuming weak system--ancilla coupling. Later in \cref{sec:cooling-interacting}, we will demonstrate that our framework here also performs well for models which cannot be exactly decomposed into two-level systems.

We first establish the two-level reduction and derive the corresponding rate equation. We then use the Landau--Zener approximation to obtain an optimal matching condition between $f_\tau(t)$ and $\Delta_{\mathrm A}(t)$, reducing the control problem to the optimization of a single coupling function. Finally, we analyze the resulting cooling errors and construct control functions with the desired scaling.

\subsection{Exact two-level reduction}

The simplest class of systems (with properly designed bath couplings) allowing an exact two-level reduction are the bipartite lattice models with hopping between different sublattices only, which we will use here as an illustration. These models have the so-called chiral symmetry as explained below. In Appendix~\ref{app:translationinvariant}, we present more examples of systems allowing an exact two-level reduction.

Specifically, we consider a bipartite free-fermion system with Hamiltonian
\begin{equation}
\hat H_{\mathrm s}
=
\sum_{i,j}
\left(
T_{ij}\hat c_{A,i}^{\dagger}\hat c_{B,j}
+\mathrm{H.c.}
\right)=\hat{\boldsymbol c}^{\dagger} H_s \hat{\boldsymbol c},
\end{equation}
where $A$ and $B$ denote the two sublattices, $i,j$ denote the sites, and $\hat{\boldsymbol c}=(\cdots,\hat c_{A,i},\cdots,\hat c_{B,i},\cdots)$ is the fermion annihilation operator basis. We assume each sublattice has $L/2$ sites, so the system contains $L$ sites in total. The single-particle Hamiltonian $H_{\mathrm s}$ is block off-diagonal and satisfies the chiral symmetry $\mathcal C H_{\mathrm s}\mathcal C^{-1}=-H_{\mathrm s}$  with $\mathcal C=\mathrm{diag}(I_A,-I_B)$. The nonzero-energy eigenmodes therefore occur in pairs at energies $\pm E\ (E>0)$ related by chiral symmetry, which we denote as a ``chiral pair''.  We denote the lower and upper edges of the positive-energy band by $E_g$ and $E_{\max}$, respectively, and its width by $W=E_{\max}-E_g$. In the numerical simulations of this section, we use the one-dimensional (1D) Su--Schrieffer--Heeger (SSH) chain $\hat H_{\mathrm s}=t^{(s)}_1\!\sum_i(\hat c_{1,i}^{\dagger}\hat c_{2,i}+\mathrm{h.c.})
+t^{(s)}_2\!\sum_i(\hat c_{1,i+1}^{\dagger}\hat c_{2,i}+\mathrm{h.c.})$, which corresponds to $T_{ij}=t_1^{(s)}\delta_{ij}+t_2^{(s)}\delta_{i,j+1}$.

We consider cooling towards the half-filled ground state $|\mathrm{GS}\rangle$ of the system, where all single-particle levels $E<0$ ($E>0$) are occupied (empty). For the cooling construction, we attach one sink ancilla $\hat s_i$ and one source ancilla $\hat r_i$ to each site of sublattice $A$. The ancilla Hamiltonian is given by Eq.~\eqref{eq:dynamical_ancilla_hamiltonian}, and the coupling is $\hat H_{\mathrm c}(t)=g_0f_\tau(t)\sum_i[\hat c_{A,i}^\dagger(\hat s_i+\hat r_i)+\mathrm{h.c.}]$ \footnote{
This coupling makes any zero-energy edge mode supported entirely on the $B$ sublattice dark to the cooling dynamics. We therefore use this construction primarily as a simple model for bulk cooling and neglect the contribution of such zero edge modes to the infidelity.
}. This design preserves the chiral symmetry for the combined system--ancilla single-particle Hamiltonian, which can therefore be written as a direct sum of different chiral pairs $H_{\rm t}(t)=H_{\mathrm s}+H_{\mathrm a}(t)+H_{\rm c}(t)
=\bigoplus_{E_n>0} H_{E_n}(t).$ For each chiral pair with energies $\pm E$ ($E>0$), the four-dimensional Hamiltonian can be decomposed as $H_E(t)
=
H_E^+(t)\otimes I
+
I\otimes H_E^-(t),$ where the two-level Hamiltonian is (see Appendix~\ref{app:quadratic_reduction})
\begin{gather}
H_E^\pm(t)
=
\frac{1}{2}\begin{pmatrix}
E\pm\Delta_{\mathrm A}(t)
&
{\sqrt{2}}{g_0f_{\tau}(t)}
\\
{\sqrt{2}}{g_0f_{\tau}(t)}
&
-E\mp\Delta_{\mathrm A}(t)
\end{pmatrix}.
\label{eq:exact_two_level_factors}
\end{gather}

The two sectors have different roles in the cooling process. The $H_E^-(t)$ sector undergoes a resonance when $\Delta_{\mathrm A}(t_E)=E$, and thus transfers excitations at energy $E$ from the system to the ancilla given the chosen ancilla initial state below \cref{eq:dynamical_ancilla_hamiltonian}. By contrast, the $H_E^+(t)$ sector with $E+\Delta_{\mathrm A}(t)>0$ never encounters a crossing in a cycle, so this sector mainly contributes to undesired nonadiabatic heating. Efficient cooling therefore requires nearly complete transfer through the resonant $H_E^-$ sector while simultaneously suppressing the nonadiabatic error in $H_E^+(t)$.

The corresponding one-cycle propagator further factorizes as $U_E=S^+(E)\otimes S^-(E)$, where $S^\pm (E)$ are $SU$$(2)$ matrices
\begin{equation}
 S^\pm(E)=
\begin{pmatrix}
\alpha_\pm(E)&\beta_\pm(E)\\
-\beta_\pm(E)^*&\alpha_\pm(E)^*
\end{pmatrix}.
\label{eq:factorized_four_level_propagator}
\end{equation}
The relation $H_E^-(t)=\sigma_xH_{-E}^+(t)\sigma_x$ implies $S^-(E)=\sigma_xS^+(-E)\sigma_x$.  The transition probabilities in the two sectors are therefore related by $|\beta_+(E)|^2=|\beta_-(-E)|^2$. Both processes can thus be described by a single function,
\begin{equation}
P_\tau(E)=|\beta_-(E)|^2\ .
\end{equation}
For $E>0$, the above $P_\tau(E)$ is the probability of heat transfer from system to ancilla in the cooling sector, while $P_\tau(-E)=|\beta_+(E)|^2$ is the probability of an undesired transition in the heating sector.

\subsection{Rate equation and optimization objective}

For ground-state preparation, the ideal cooling, therefore, requires $P_\tau(E)=1$ and $P_\tau(-E)=0$ for $E>0$ within the system's single-particle band. Namely, the ideal $P_\tau(E)$ is the step function $\Theta(E)$, as shown in Fig.~\ref{fig:cooling_protocol}(d). 
For simplicity, we quantify deviations from ideal cooling over the system spectrum $\mathcal S=[-E_{\max},-E_g]\cup[E_g,E_{\max}]$
by the stopband and passband errors,
\begin{align}\label{Eq:stopbandpassbanderror}
\mathcal A_{\rm stop}(\tau;\mathcal S)
&=\int_{E_g}^{E_{\max}} P_\tau(-E)\,dE,\\
\mathcal A_{\rm pass}(\tau;\mathcal S)
&=\int_{E_g}^{E_{\max}} [1-P_\tau(E)]\,dE,
\end{align}
giving the total error as $\mathcal A(\tau;\mathcal S)=\mathcal A_{\rm stop}(\tau;\mathcal S)+ \mathcal A_{\rm pass}(\tau;\mathcal S)$. 
The interval $[-E_g,E_g]$ is denoted as the transition band, which vanishes in the gapless case. The passband error mainly controls the cooling rate at which entropy is transferred to the ancilla, whereas the stopband error mainly determines the steady-state error arising from non-adiabatic heating. Suppressing both errors can be limited by the finite energy resolution $\delta E \gtrsim 1/\tau$. We therefore study how the total transfer error $\mathcal A(\tau;\mathcal S)$ scales with the cycle period $\tau$.

To determine how the transfer errors in Eq.~\eqref{Eq:stopbandpassbanderror} affect the cooling dynamics, we consider the excitation density of each chiral pair of energy levels $\pm E$
\begin{equation}
p_E
=
\frac{1}{2}
\left(
n_{+E}
+
1-n_{-E}
\right),
\qquad E>0.
\label{eq:pair_excitation_density}
\end{equation}
where $n_{\pm E}
\equiv
\left\langle
\hat c_{\pm E}^\dagger
\hat c_{\pm E}
\right\rangle$ is the fermion number in level $\pm E$ of the system. The goal of ground state preparation is to decrease $p_E$ towards $0$.  Using the propagator in Eq.~\eqref{eq:factorized_four_level_propagator}, one arrives at the exact rate equation (see Appendix~\ref{app:quadratic_reduction})
\begin{equation}
p_E^{(m+1)}
=
[1-P_\tau(E)]p_E^{(m)}
+
P_\tau(-E)[1-p_E^{(m)}].
\label{eq:pair_rate_equation}
\end{equation}
The first term characterizes excitations that remain in the system due to incomplete heat transfer to ancillas, while the second term describes excitations generated by heating.

The dependence on the initial excitation density is controlled by
\begin{equation}
\lambda_E
=
1-P_\tau(E)-P_\tau(-E).
\label{Eq:single_cycle_condition}
\end{equation}

For $|\lambda_E|<1$, the excitation density approaches
\begin{equation}
p_E^\infty
=
\frac{P_\tau(-E)}
{P_\tau(E)+P_\tau(-E)}= \frac{P_\tau(-E)}
{1-\lambda_E}\ ,
\label{Eq:steadystaterate}
\end{equation}
and the general solution is $p_E^{(m)}-p_E^\infty
=\lambda_E^m\left(p_E^{(0)}-p_E^\infty\right).$ 
We emphasize that Eqs.~\eqref{eq:pair_rate_equation}-\eqref{Eq:steadystaterate} are exact for the cooling setup considered here, as shown by the comparison with numerical results in Fig.~\ref{fig:filter_error_scaling}(a). For $|\lambda_E|\ll1$, the steady-state excitation density satisfies $p_E^\infty\simeq P_\tau(-E)$. 

In particular, we refer to $\lambda_E=0$ as the ideal single-cycle condition, which gives $p_E^{(1)}=p_E^\infty=P_\tau(-E)$ independent of $p_E^{(0)}$, i.e., the steady state is prepared in a single cycle. However, the exact condition $\lambda_E=0$ is not required for practical single-cycle cooling. With $|p_E^{(0)}-p_E^\infty|\leq1$, we have $|p_E^{(1)}-p_E^\infty|\leq |\lambda_E|$. Therefore, for an energy $E$ with $|\lambda_E|\lesssim p_E^\infty\ll1$, one cycle readily produces an excitation density of the same order as its steady-state value $p_E^\infty$.
Fig.~\ref{fig:filter_error_scaling}(b) shows that, $|\lambda_E|$ is the largest near zero energy and strongly suppressed over most of the system band. Consistent with the condition $|\lambda_E|\lesssim p_E^\infty$, the dynamical protocols in Fig.~\ref{fig:filter_error_scaling}(a) reduce the excitation densities to values comparable to their steady-state values after one cycle.

Over the full spectral interval, the corresponding bound is
\begin{equation}
\begin{aligned}
\int_{E_g}^{E_{\max}}
|p_E^{(1)}-p_E^\infty|\,dE
&\leq
\int_{E_g}^{E_{\max}}|\lambda_E|\,dE\\
&\leq
\mathcal A_{\rm pass}+\mathcal A_{\rm stop}.
\end{aligned}
\label{eq:single_cycle_spectral_bound}
\end{equation}
Thus, the total transfer error bounds the total deviation from the steady-state excitation density after one cycle. When this deviation is at most of order $\int_{E_g}^{E_{\max}}p_E^\infty\,dE$, a single cycle already achieves an excitation error of the same order as, or even smaller than, its steady-state value.

The pair excitation densities also determine the many-body fidelity for an initial Gaussian state that factorizes in the energy eigenmode basis of the system and satisfies $n_{+E}^{(0)}+n_{-E}^{(0)}=1$ for each chiral pair.  These properties are preserved by the factorized cooling dynamics in Eq.~\eqref{eq:factorized_four_level_propagator}(see Appendix~\ref{App:manybodyfidelity}). The fidelity with the half-filled ground state is therefore
\begin{equation}
\mathcal F_m
=
\prod_{E>0}
\left(1-p_E^{(m)}\right)^2.
\label{eq:manybody_fidelity_product}
\end{equation}

Taking the logarithm gives an additive measure of the error over all chiral pairs,
\begin{equation}
-\frac{\ln\mathcal F_m}{L}
=
-\frac{2}{L}
\sum_{E>0}\ln\left(1-p_E^{(m)}\right).
\end{equation}
We use this log-fidelity density to characterize the many-body cooling performance in the free-fermion benchmarks.

When $p_E^\infty\ll1$ and $|\lambda_E|\ll1$ throughout the spectrum, the steady-state excitation density satisfies
$p_E^\infty=P_\tau(-E)/(1-\lambda_E)\simeq P_\tau(-E)$, giving 
\begin{equation}
-\frac{\ln\mathcal F_{\rm ss}}{L}
\simeq
\frac{2}{L}\sum_{E>0}P_\tau(-E)
\simeq
\frac{\mathcal A_{\rm stop}(\tau;\mathcal S)}{W}\ ,
\label{eq:steady_state_log_fidelity}
\end{equation}
where $\mathcal F_{\rm ss}=\mathcal F_{\infty}$ is the fidelity of the steady state. The last approximation assumes a flat density of states, $\rho(E)=L/(2W)$ for $E\in[E_g,E_{\max}]$. Thus, in this regime, suppressing the stopband error reduces the steady-state log-fidelity density.

The ground-state infidelity after a finite number of cycles is generically bounded by the total transfer error (see Appendix~\ref{App:manybodyfidelity}),
\begin{equation}
1-\mathcal F_m
\lesssim
\frac{L}{W}
\left[
\mathcal A_{\mathrm{pass}}(\tau)
+
\mathcal A_{\mathrm{stop}}(\tau)
\right]
=
\frac{L}{W}\mathcal A(\tau;\mathcal S).
\label{eq:finite_cycle_infidelity_bound}
\end{equation}
Here $\mathcal A_{\rm pass}$ penalizes the incomplete excitation transfer and $\mathcal A_{\rm stop}$ penalizes the generation of new excitations. These relations motivate using $\mathcal A(\tau;\mathcal S)$ as a scalar surrogate for the bi-objective cooling optimization problem. In the single-cycle cooling regime $|\lambda_E|\lesssim p_E^\infty\ll1$, minimizing $\mathcal A(\tau;\mathcal S)$ becomes the exact cooling objective.

\begin{figure}[t]
\centering
\includegraphics[width=0.9\columnwidth]{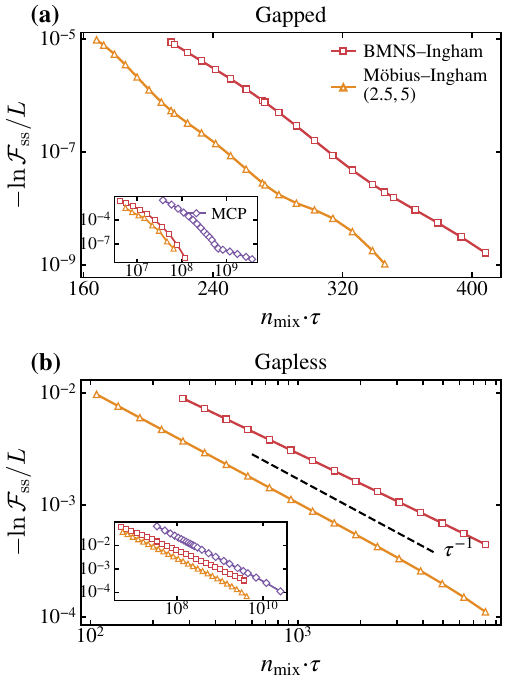}
\caption{Steady-state error and cooling time in the SSH model.
Log-fidelity density $-\ln\mathcal F_{\rm ss}/L$ versus physical cooling time $n_{\rm mix}\tau$ for (a) gapped and (b) gapless chains. Here $n_{\rm mix}$ is the first cycle for which $-\ln\mathcal F_n/L$ remains within $1\%$ of its steady-state value for three consecutive cycles. Main panels use $\Gamma_0=25$; the insets compare all three protocols at $g_0=0.005$, including static-ancilla MCP~\cite{PRXQuantum.6.010361}. The dashed line in (b) indicates $\tau^{-1}$. All other parameters are the same as in Fig.~\ref{fig:filter_error_scaling}(c,d).}
\label{fig:ssh_cooling_mixing}
\end{figure}

\begin{figure*}[t]
\centering
\includegraphics[width=\textwidth,height=\paperfigureheight,keepaspectratio]{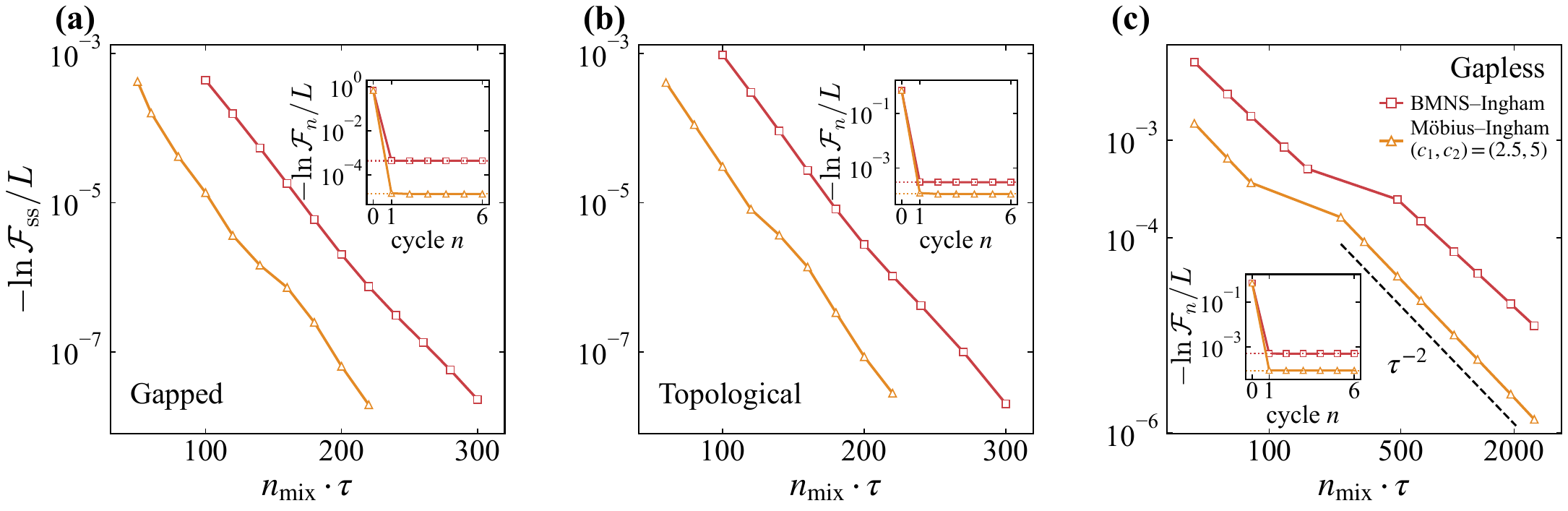}
\caption{Steady-state cooling in the Haldane model.
Log-fidelity density $-\ln\mathcal F_{\rm ss}/L$ versus cooling time $n_{\rm mix}\tau$ for (a) gapped, (b) topological, and (c) gapless cases with $m=1,\,0,\,0.3\sqrt{3}$, respectively, and $\phi=\pi/2$. The dashed line in (c) indicates $\tau^{-2}$. Insets show $-\ln\mathcal F_n/L$ versus cycle number.
We use $1000\times1000$ unit cells ($L=2\times10^6$ system sites), periodic boundaries, and $C_0=I/2$. Each system site couples to an empty sink and a filled source, giving
$L_a=2L$ ancilla modes in total.
We use $g_0=3$, $\alpha=1/3$. The mixing criterion is the same as in
Fig.~\ref{fig:ssh_cooling_mixing}.
}
\label{fig:twodim_cooling_mixing}
\end{figure*}

\subsection{Matching condition and error scalings}
\label{subsec:asymptotic_design}

Minimizing the total transfer error $\mathcal A(\tau;\mathcal S)$ is equivalent to approximating the ideal step function shown in Fig.~\ref{fig:cooling_protocol}(d). This requires efficient excitation transfer in the cooling sector and suppressed nonadiabatic heating in the heating sector. Smooth protocols with a single Landau--Zener-type crossing provide a simple physical picture. A monotonic ancilla energy ramp brings each positive-energy mode through resonance once, where sufficiently strong coupling allows nearly complete excitation transfer. By increasing the cycle period $\tau$, we further suppress the nonadiabatic error in the heating sector. Even simple sine and sine-squared controls can therefore reduce the cooling error shown in Fig.~\ref{fig:filter_error_scaling}(c,d).

To achieve faster error decay with the cycle period $\tau$, in the following, we will optimize the coupling function $f_\tau(t)$ and ramping function $\Delta_{\mathrm A}(t)$. We first derive an optimal matching condition between the ancilla ramping function and the coupling function to suppress the passband error. This condition determines the ramp from the coupling function. We then analyze the stopband error scaling in gapless and gapped spectra to identify how the coupling function should be chosen.

We summarize these results in four rules for designing protocols with optimal error scaling:
\begin{enumerate}
    \item The ancilla ramp $\Delta_A(t)$ should match the coupling $f_\tau(t)$ via Eq.~\ref{eq:optimal_matching_condition}. 
    \item The coupling function $f_\tau(t)$  should be smooth, and supported on the interval $[0,\tau].$
    \item The Fourier tails of $\tilde f_\tau(\omega)$ should decay rapidly for $|\omega| \gg E_g$.
    \item For gapless cooling, the rescaled function $\tau f_\tau(\tau x)$ approaches a $\tau$-independent function at large $\tau$.
\end{enumerate}

\subsubsection{Optimal matching condition via Landau--Zener approximation}
\label{subsubsec:LZ}
To reach the steady state within $O(1)$ cycles, we need to suppress the passband error $\mathcal A_{\rm pass}$. We therefore analyze excitation transfer in the resonant sector $H_E^-(t)$ within the Landau--Zener picture.

For $0<E<\Delta_0$, the crossing occurs at $t_E$ defined by $\Delta_{\mathrm A}(t_E)=E$.  Approximating the ramping function as linear and the coupling as constant near $t_E$ gives the Landau--Zener transfer probability~\cite{zener1932non},
\begin{equation}
P_\tau(E)
\simeq
1-e^{-\Gamma_E},
\,
\Gamma_E
=
\frac{2\pi g_0^2|V|^2|f_\tau(t_E)|^2}
{-\dot{\Delta}_{\mathrm A}(t_E)}\ ,
\end{equation}
where $\dot{\Delta}_{\mathrm A}(t)$ is the time derivative of $\Delta_{\mathrm A}(t)$, and $V$ is the coupling matrix element ($V=1/\sqrt{2}$ in the example of \cref{eq:exact_two_level_factors}). The estimated passband error is therefore controlled by the ratio between the squared coupling and the ramping speed at resonance. We note that for the Landau--Zener approximation to hold, $f_\tau(t)$ should be smooth and slowly varying. Additionally, since a smooth time-dependent phase of $f_\tau(t)$ can be absorbed into the ancilla Hamiltonian $\Delta_{\mathrm A}(t)$ by a rotating-frame transformation, we take $f_\tau(t)$ to be real without loss of generality.

For a given coupling function, we optimize over monotonically decreasing ramps from $\Delta_0$ to zero. Within the Landau--Zener approximation, minimizing the largest passband error is equivalent to maximizing the minimum of the transfer exponent $\Gamma_E$, which gives an optimal matching condition
\begin{equation}
-\dot{\Delta}_{\mathrm A}(t)
=
\Delta_0 |f_{\tau}(t)|^2,
\label{eq:optimal_matching_condition}
\end{equation}
as derived in Appendix~\ref{app:optimal_matching}.

When the condition in \cref{eq:optimal_matching_condition} is achieved, the transfer exponent becomes independent of the resonance energy:
\begin{equation}
\Gamma_E=\Gamma_0,
\qquad
\Gamma_0
=
\frac{2\pi g_0^2|V|^2}{\Delta_0}.
\end{equation}
For strong couplings with $\Gamma_0\gg1$, the Landau--Zener approximation therefore gives nearly complete excitation transfer. This matching condition is related to offset-independent adiabaticity in nuclear magnetic resonance~\cite{1996JMagR.120..133T,GARWOOD2001155,optimalrelation2002,NMRSilverPRA}.

 Once the condition in Eq.~\eqref{eq:optimal_matching_condition} is imposed, the coupling function $f_\tau(t)$ and $\Delta_0$ determine the ancilla ramp $\Delta_{\mathrm A}(t)$, with examples illustrated in Fig.~\ref{fig:cooling_protocol}(b,c).

The Landau--Zener approximation gives a passband error $\mathcal A_{\rm pass}\simeq We^{-\Gamma_0}$, which holds in the large cycle period $\tau$ limit(see Appendix~\ref{App:singlecycleerror}). However, the finite-period passband error $\mathcal A_{\rm pass}(\tau;\mathcal S)$ further depends on the control functions and the evolution away from the main resonance.
Although a monotonic ancilla ramp gives a single resonance at $\Delta_{\mathrm A}(t)=E$, the Fourier components $\widetilde{f}_\tau(\omega)$ of the coupling $f_\tau(t)$ at nonzero frequencies $\omega\neq0$ can introduce additional resonances at $\Delta_{\mathrm A}(t)=E\pm\omega$ via the rotating frame transformation. These additional resonances can typically transfer excitations back to the system, increasing the passband error.

As a heuristic understanding, consider two incoherent resonances with the same transfer probability $P_{\tau}(E)$ happening at distinct times in a cycle. The net transfer probability $P_{\tau}^{(2)}(E)$ would be~\cite{Landau-Zener-Stinterferometry} 
\begin{equation}\label{eq:two-resonance}
P_{\tau}^{(2)}(E)
=
2P_{\tau}(E)
[1-P_{\tau}(E)]\ .
\end{equation}
In the strong-coupling regime with $P_\tau(E)>1/2$, one has $P_{\tau}^{(2)}(E)<P_{\tau}(E)$, namely, two resonances are less efficient than one resonance.  Therefore, the Fourier tail of $\widetilde{f}_\tau(\omega)$ at nonzero $\omega$ can increase the passband error $\mathcal A_{\rm pass}$ by transferring excitations back into the system after the main resonance.

As discussed in the next section, the coupling's Fourier tail $\widetilde{f}_\tau(\omega)$ also contributes to nonadiabatic heating in the stopband. The common error source suggests that, for a finite cycle period $\tau$ satisfying $We^{-\Gamma_0}\ll\mathcal A_{\rm stop}(\tau;\mathcal S)\ll W$, one has $\mathcal A_{\rm pass}(\tau;\mathcal S)=  O\!\left(\mathcal A_{\rm stop}(\tau;\mathcal S)\right)$, which is verified in Appendix~\ref{App:singlecycleerror}. In this regime, we have  $\mathcal A(\tau;\mathcal S)=O(\mathcal A_{\rm stop}(\tau;\mathcal S))$. We therefore turn to optimizing the coupling function to suppress the stopband error $\mathcal A_{\rm stop}$.

\subsubsection{Optimal stopband error scalings}

With the optimal matching condition of \cref{eq:optimal_matching_condition} imposed, we now optimize the coupling function $f_\tau(t)$ to suppress the stopband error $\mathcal A_{\rm stop}$. Since whether a system is gapped or gapless is often unknown a priori, our goal is to find a function $f_\tau(t)$ that is as optimal as possible for gapped and gapless systems simultaneously. In the following, we analyze the two cases separately.

\paragraph{Gapped spectra: suppression of Fourier tails.}
\label{subsubsec:gapped_scaling}

For systems with gapped spectra ($E_g>0$), the heating is governed by the heating sector $H_E^+(t)$, and energies $|E|<E_g$ do not contribute to the stopband error. 

To analyze the period $\tau$ dependence of the stopband error, we introduce a dimensionless time $x=t/\tau\in[0,1]$. For now we assume the coupling function for different $\tau$ takes the form
\begin{equation}\label{eq:ft-sx}
f_\tau(t)=\tau^{-1/2}s(t/\tau)\ ,
\end{equation}
where $s(x)$ is a dimensionless real function normalized to $\int_0^1 |s(x)|^2dx=1$.

A first source of error comes from truncating the coupling function $f_\tau(t)$ to the interval $[0,\tau]$, which can produce algebraic Fourier tails that depend on the endpoint smoothness of $s(x)$.

Specifically, suppose the first non-vanishing endpoint derivative of $s(x)$ is of order $m\geq0$, i.e., the $k$-th derivative $s^{(k)}(0)=s^{(k)}(1)=0$ for $k=0,\ldots,m-1$, while at least one of $s^{(m)}(0)$ and $s^{(m)}(1)$ is nonzero. As shown in Appendix~\ref{app:gapped_switching}, by first-order adiabatic perturbation theory and repeated integration by parts, one finds
\begin{equation}
\mathcal A_{\rm stop}^{(1)}(\tau;S_{\rm gapped})
=
O\!\left( \tau^{-(2m+1)}\right),
\label{eq:gapped_error_powerlaw}
\end{equation}
where the superscript $(1)$ denotes the first-order estimate. For example, the sine function $s(x)\propto\sin(\pi x)$ and sine-squared function $s(x)\propto\sin^2(\pi x)$ have $m=1$ and $m=2$, respectively, giving the $\tau^{-3}$ and $\tau^{-5}$ scaling observed in Fig.~\ref{fig:filter_error_scaling}(d). Intuitively, this is because $f_\tau(t)$ is not smooth at $t=\tau$, with $f_\tau(t)\propto \tau^{-(2m+1)/2}(\tau-t)^m$ as $t\to\tau^-$ and $f_\tau(t)=0$ for $t>\tau$ (and similarly for $t=0$), which yields a dominant heating probability to the leading order proportional to $g_0^2|f_\tau(E^{-1})|^2/E^2\sim g_0^2E_g^{-2(m+1)}\tau^{-(2m+1)}$.

At large $\tau$, the error is smaller for higher $m$. This motivates choosing a function $s(x)$ that has all orders of derivatives $s^{(m)}(0)=s^{(m)}(1)=0$ flat at the end points, making it smooth up to infinite order. This does not yet fix the choice of $s(x)$. As we argued above in \cref{eq:two-resonance}, the remaining error can be further suppressed by choosing $s(x)$ with a rapidly decaying Fourier transform $\widetilde s(k)$ (see Appendix~\ref{app:gapped_switching}).

However, for a dimensionless coupling function $s(x)$ that is nonzero only in the interval $[0,1]$ and independent of $\tau$, the Paley–Wiener theorem rules out strict exponential decay of its Fourier tails $\widetilde s(k)$ as $|k|\to\infty$~\cite{paley1934fourier}~\footnote{Coupling functions optimized at each $\tau$, with a general form that depends on the cycle period, may circumvent this limitation.}. We therefore focus on coupling functions with almost-exponentially decaying Fourier tails, as constructed in Sec.~\ref{subsec:mobius_ingham_design}.

\paragraph{Gapless spectra: suppression of transition region error.}
\label{subsubsec:gapless_scaling}

For systems with gapless spectra, $E_g=0$, and the transition near $E=0$ dominates the stopband error. 
Heuristically, a protocol of period $\tau$ has a finite energy resolution of order $1/\tau$. Since $P_\tau(E)$ necessarily changes by an amount of $O(1)$ around $E=0$, the stopband error in \cref{Eq:stopbandpassbanderror} for a system with constant single-particle density of states at $E=0$ will scale as
\begin{align}\label{eq:gapless-Astop-bound}
\mathcal A_{\rm stop}(\tau;\mathcal S_{\rm gapless})
\gtrsim \frac{1}{\tau}.
\end{align}
Saturation of the above bound holds when the two-level Schr\"odinger equation is independent of $\tau$ at fixed $\tau E$.  At fixed coupling strength $g_0$, a sufficient condition is that the rescaled control function  $\tau f_\tau(\tau x)$ approaches a $\tau$-independent function at large $\tau$ (see Appendix~\ref{Appendix:Gaplessscaling}). In \cref{subsec:mobius_ingham_design}, we design the M\"obius--Ingham family functions to satisfy this scaling condition and thus saturate the bound in \cref{eq:gapless-Astop-bound}.

Physically, this error is governed by the number of energy modes within the unresolved window $|E|\lesssim1/\tau$. For a density of states $\rho(E)\propto|E|^\nu$ with $\nu>-1$, the same argument gives
\begin{equation}
\mathcal A_{\rm stop}(\tau;\mathcal S_{\rm gapless,\rho})
\sim \tau^{-(\nu+1)}.
\label{eq:gapless_dos_scaling}
\end{equation}

\subsection{M\"obius--Ingham control functions}
\label{subsec:mobius_ingham_design}

We now construct coupling functions that satisfy the design rules identified above. For gapped spectra, our analysis above requires us to find a function $s(x)$ supported on $[0,1]$ that is flat to all orders of derivatives at both endpoints and has rapidly decaying Fourier tails. Although compact support rules out strictly exponential Fourier decay, the Ingham functions can achieve almost-exponential decay and thus provide a near-optimal starting point. As a concrete choice, we can use the BMNS--Ingham function constructed in Ref.~\cite{Bachmanningham}, which we denote as $s_{\mathrm I}(x)$ (see Appendix~\ref{App:inghamfunction} for its construction).

However, the function $f_\tau(t)$ given by \cref{eq:ft-sx} with $s(x)$ being the BMNS-Ingham function $s_{\mathrm I}(x)$ does not satisfy the scaling condition for gapless cooling. To achieve optimal gapless scaling as well, we introduce a period-dependent M\"obius transformation of $s_{\mathrm I}(x)$. The resulting M\"obius--Ingham functions satisfy this condition while preserving endpoint flatness and almost-exponential Fourier decay.

Starting from the normalized BMNS--Ingham function $s_{\mathrm I}(x)$, we define the integral of its square,
\begin{equation}
A(x)=\int_0^x s_{\mathrm I}(u)^2\,du.
\end{equation}
Applying the M\"obius transformation to $A(x)$, and imposing the optimal matching condition \cref{eq:optimal_matching_condition}, the resulting control functions we propose are
\begin{align}\label{Eq:mobiusprotocol}
\Delta_{\rm A}(\tau x)
&=
\Delta_0
\frac{1-A(x)}
{1+(\Delta_0\tau-1)A(x)},
\\
f_\tau(\tau x)
&=
\frac{\sqrt{\Delta_0}\,s_{\mathrm I}(x)}
{1+(\Delta_0\tau-1)A(x)}\ ,\label{Eq:mobiusprotocol2}
\end{align}
which we refer to as a M\"obius--Ingham control.

For gapped spectra, the transformed coupling function remains flat to all orders at both endpoints. Additionally, the Fourier analysis in Appendix~\ref{subsec:reweighted_mobius_ingham_class} still gives an almost-exponentially suppressed first-order stopband error bound for $E_g\tau\gg1$:
\begin{equation}
\mathcal A_{\rm stop}^{(1)}
(\tau;\mathcal S_{\rm gapped})
\leq
CW\exp\!\left[
-a_1\frac{E_g\tau}{\ln^3(E_g\tau)}
\right],
\end{equation}
where $C,a_1>0$ are constants independent of $\tau$. In fact, simple transformations of the Ingham class functions always preserve the infinite-order flatness at endpoints and the almost-exponential Fourier decay property~\cite{
Compositioninultradifferentiableclasse,ultradifferentiablefunctionclass}.

Neglecting the higher-order nonadiabatic corrections, and since the passband and stopband errors are of the same order in the regime $We^{-\Gamma_0}\ll\mathcal A_{\rm stop}\ll W$, we have for gapped systems
\begin{equation}\label{Eq:mobiusgappederror}
\mathcal A(\tau;\mathcal S_{\rm gapped})
=O\!\left(
\mathcal A_{\rm stop}^{(1)}
(\tau;\mathcal S_{\rm gapped})\right )
\end{equation}
 for  $1\ll E_g\tau\lesssim (\Gamma_0/a_1)\ln^3(\Gamma_0/a_1)$ with $\Gamma_0/a_1\gg1$. Notice that one can increase the coupling strength $\Gamma_0$ to further reduce the Landau--Zener passband error $We^{-\Gamma_0}$.  At the precision of interest, however, a fixed $\Gamma_0\gg1$ already provides effective cooling, while allowing $\Gamma_0$ to grow with the cycle period does not necessarily improve the performance (see Appendix~\ref{App:singlecycleerror}). Therefore, jointly optimizing $\Gamma_0(\tau)$ and the coupling function $f_\tau(t)$ is left for future work, while in this work we focus on fixed $\Gamma_0\gg 1$.

For gapless spectra, the denominator in Eq.~\eqref{Eq:mobiusprotocol} grows linearly with $\tau$, so at large $\tau$, the rescaled coupling function $\tau f_\tau(\tau x)$ approaches a $\tau$-independent function. Therefore, the corresponding large-$\tau$ limit saturates the optimal scaling for gapless systems:
\begin{equation}
\mathcal A(\tau;\mathcal S_{\rm gapless})
\simeq
\frac{a_2}{\tau},
\end{equation}
given that the transition error dominates over the almost-exponential decay error in Eq.~\eqref{Eq:mobiusgappederror} for the regime $E\gg 1/\tau$, namely, $a_2/\tau\gg \max\{\mathcal A(\tau;\mathcal S_{\rm gapped}), We^{-\Gamma_0}\}$.

The scaling coefficients $(a_1,a_2)$ can be further optimized in our M\"obius--Ingham control without changing the scaling laws. We consider a simple two-parameter deformation $s_{\mathbf c}(x)=s_{\mathrm I}(x)\exp\!\left[c_1(2x-1)+c_2(2x-1)^2\right]$ with $\mathbf c=(c_1,c_2)$ and apply the same construction using the corresponding $Z_{\mathbf c}$ and $A_{\mathbf c}(x)$. This analytic weight preserves the endpoint flatness and the scaling laws while changing the error coefficients $(a_1,a_2)$.

We scan $\mathbf c=(c_1,c_2)$ to determine the trade-off between the gapped decay coefficient $ a_1$ and the gapless coefficient $a_2$. The optimal $\mathbf c=(c_1,c_2)$ gives the Pareto frontier within this family (see Appendix~\ref{subsec:reweighted_mobius_ingham_class}). Hereafter, we set $(c_1,c_2)=(2.5,5)$ for our M\"obius--Ingham control, as a representative choice near this frontier.

We test the resulting protocol through both single-cycle and steady-state cooling in the SSH chain. In Fig.~\ref{fig:filter_error_scaling}(c)and \ref{fig:filter_error_scaling}(d), we show the single-cycle log-fidelity density versus $\tau$ for gapless and gapped chains, respectively. The error for gapless spectra follows the predicted $1/\tau$ scaling, while the error for gapped spectra decreases rapidly, consistent with the almost-exponential suppression.

Similar behavior appears in the steady-state results in Fig.~\ref{fig:ssh_cooling_mixing}, where we compare the log-fidelity density with the physical cooling time $n_{\rm mix}\tau$, where $n_{\rm mix}$ denotes the number of cooling cycles needed for the cooling error to reach its steady-state value up to a relative error $\epsilon$. The insets extend the comparison to the weak-coupling regime and include the MCP protocol of Ref.~\cite{PRXQuantum.6.010361}, showing that the M\"obius--Ingham protocol indeed provides a better trade-off between steady-state error and cooling time.

\begin{figure}[t]
\centering
\includegraphics[height=\paperfigureheight]{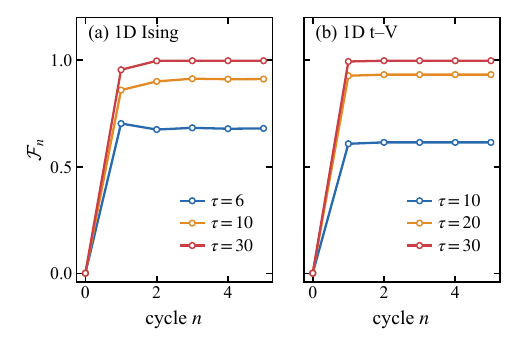}
\caption{Cooling in interacting one-dimensional chains.
Ground-state fidelity $\mathcal F_n$ versus cycle number under M\"obius--Ingham cooling for (a) the Ising chain with $J=h_x=1$, $h_z=0.2$, and periodic boundaries, and (b) the $t$--$V$ chain with $t=1$, $V=2$,
and open boundaries(OBC). Both models have $L=8$ system sites and $L_a=L$ ancillas, with $g_0=10$ and $(c_1,c_2)=(2.5,5)$. The initial ancilla energies are set by $\Delta_0\simeq 22.08$ in (a) and $10.40$ in (b). Initial system states are Haar-random pure states in the full Hilbert space.}
\label{fig:interacting_reset_fidelity}
\end{figure}

\section{Cooling generic many-body systems: free fermions and beyond}
\label{sec:cooling-interacting}

In the following, we will test the performance of our protocols in \cref{sec:cooling-mechanism} in two-dimensional (2D) free-fermion systems, interacting 1D and 2D models. Across these models, the M\"obius--Ingham protocol can approach the steady state within only one or two cycles, demonstrating its effectiveness beyond the analytically tractable settings considered above.

\subsection{Two-Dimensional free-fermion model }\label{Sec:freenumerics}

A natural generalization is to examine free fermions in higher dimension. Here we consider the Haldane model on a honeycomb lattice at half filling~\cite{PhysRevLett.61.2015}, which allows us to compare cooling in trivial, topological, and gapless bands:
\begin{align}
\hat H_{\mathrm s}
&=
\sum_{\langle i,j\rangle}
\left(\hat c_i^\dagger\hat c_j+\mathrm{h.c.}\right)
+
0.1\sum_{\langle\langle i,j\rangle\rangle}
\left(e^{i\nu_{ij}\phi}
\hat c_i^\dagger\hat c_j+\mathrm{h.c.}\right)
\nonumber\\
&\quad+
m\sum_i\xi_i\hat c_i^\dagger\hat c_i,
\label{eq:Hhaldane}
\end{align}
where $\nu_{ij}=\pm1$ specifies the orientation-dependent hopping with $\nu_{ji}=-\nu_{ij}$,  and $\xi_i=+1$ ($-1$) on sublattice $A$ ($B$) gives a staggered on-site potential. Each system site couples to one sink and one source ancilla.

Fig.~\ref{fig:twodim_cooling_mixing} compares the steady-state log-fidelity density as a function of the cooling time $n_{\rm mix}\tau$. Both gapped phases exhibit rapid error suppression consistent with almost-exponential decay. The gapless case instead follows the $\tau^{-2}$ scaling expected from the linear density of states near the Dirac point, $\rho(E)\propto|E|$, in accordance with Eq.~\eqref{eq:gapless_dos_scaling}. The insets of Fig.~\ref{fig:twodim_cooling_mixing} further show that the log-fidelity density after a single cycle is already comparable to its steady-state value.

\subsection{Interacting systems}
\label{Sec:interacting}
The resonant-transfer mechanism in \cref{sec:cooling-mechanism} also applies to interacting systems. When the energy released by a system transition matches the energy required to excite the ancilla, the coupling can transfer the system to the lower-energy state while exciting the ancilla. Near an isolated resonance, this process is approximately described by a two-level Hamiltonian of the form $H_E^-(t)$. We therefore test whether the same control functions also enable efficient cooling in interacting many-body systems, employing numerical exact diagonalization.

\subsubsection{One-dimensional Ising and \texorpdfstring{$t$--$V$}{t--V} models}
\label{Sec:1Dinteracting}

In 1D, we consider two models. The first model is an Ising chain with transverse and longitudinal fields,
\begin{equation}
\hat H_{\rm s}^{\rm Ising}
=
-J\sum_{j=1}^{L}\sigma_j^z\sigma_{j+1}^z
-h_x\sum_{j=1}^{L}\sigma_j^x
-h_z\sum_{j=1}^{L}\sigma_j^z,
\end{equation}
where the nonzero longitudinal field $h_z$ breaks integrability. The Ising chain couples locally to spin ancillas through $Y$--$Y$ interactions.

The second model we consider is a spinless-fermion $t$--$V$ chain at half filling,
\begin{align}
\hat H_{\rm s}^{\rm tV}
&=
-\sum_{j=1}^{L-1}
\left(\hat c_j^\dagger\hat c_{j+1}
+\mathrm{h.c.}\right)
\nonumber\\
&\quad+
V\sum_{j=1}^{L-1}
\left(\hat n_j-\frac12\right)
\left(\hat n_{j+1}-\frac12\right).
\end{align}
For the $t$--$V$ chain, each even system site couples through hopping to an empty sink fermion ancilla and a filled source fermion ancilla.

Fig.~\ref{fig:interacting_reset_fidelity} shows the ground-state fidelity under repeated M\"obius--Ingham cooling cycles. In both models,  the M\"obius--Ingham protocol reaches a single-cycle fidelity close to its steady-state value. These results show that the control functions derived from the free-fermion analysis in \cref{sec:cooling-mechanism} also enable rapid cooling in the interacting models.

\subsubsection{Two-dimensional Fermi--Hubbard model}
\label{Sec:2Dfhnumerics}

Finally, we consider the half-filled two-dimensional Fermi--Hubbard model~\cite{Xu2025,PhysRevLett.129.123201},
\begin{equation}
\hat H_{\rm s}
=
-J\sum_{\langle i,j\rangle,\sigma}
\left(
\hat c_{i\sigma}^\dagger\hat c_{j\sigma}
+\mathrm{h.c.}
\right)
+
U\sum_i
\hat n_{i\uparrow}\hat n_{i\downarrow}.
\label{eq:2d_fermi_hubbard}
\end{equation}
Here we use a different coupling geometry that preserves the system particle number.
We introduce four ancilla layers and couple system and ancilla hopping on the same bond:
\begin{equation}
\hat H_c(t)
=
g_0 f_{\tau}(t)\sum_{\langle i,j\rangle}
\left[
\sum_\sigma
\left(
\hat c_{i\sigma}^\dagger\hat c_{j\sigma}
+\mathrm{h.c.}
\right)
\right]
\sum_{\ell=1}^{4}
\left(
\hat d_{i\ell}^\dagger\hat d_{j\ell}
+\mathrm{h.c.}
\right).
\label{eq:fh_ancilla_coupling}
\end{equation}
Each ancilla layer has staggered single-particle energies $+\Delta_{\rm A}(t)$ and $-\Delta_{\rm A}(t)$, following Eq.~\eqref{eq:dynamical_ancilla_hamiltonian}.

\begin{figure}[t]
\centering
\includegraphics[height=\paperfigureheight]{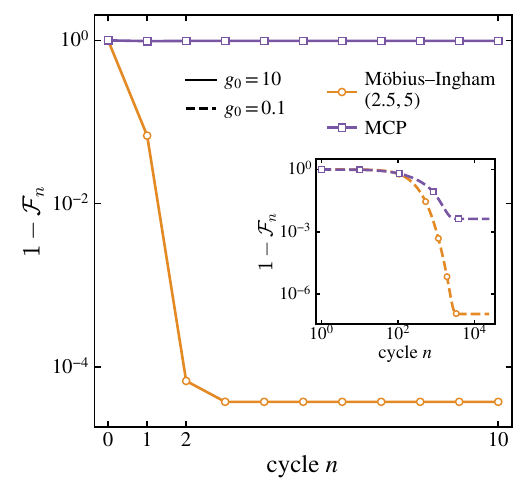}
\caption{Cooling of the half-filled $2\times3$ Fermi--Hubbard model with $L_a=24$ ancilla modes.
Ground-state infidelity $1-\mathcal F_n$ versus cycle number for M\"obius--Ingham and MCP. The main panel has $g_0=10$, and the inset has
$g_0=0.1$, with cycle period $\tau=80$ for both protocols in both cases.
M\"obius--Ingham uses $\Delta_0=2.8$, while MCP uses static ancillas. The parameters are $J=1,U=8$  and OBC.}
\label{fig:fh_half_filling_cooling}

\end{figure}

In Fig.~\ref{fig:fh_half_filling_cooling}, we compare M\"obius--Ingham cooling with the static-ancilla modulated-coupling protocol (MCP) of Ref.~\cite{PRXQuantum.6.010361}. At strong coupling, the M\"obius--Ingham protocol
reduces the infidelity below $10^{-4}$ within two cycles, whereas MCP produces little cooling at the same cycle period and coupling strength.  The inset shows that M\"obius--Ingham also achieves a better error--time trade-off at weak coupling.

\begin{figure}[t]
\centering
\includegraphics[height=\paperfigureheight]{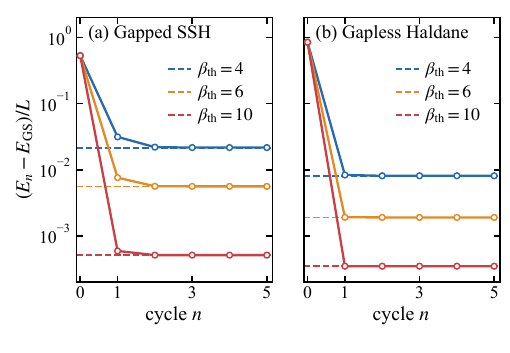}
\caption{Energy density $(E_n-E_{\mathrm{GS}})/L$ versus
cycle $n$ for DK cooling in (a) a gapped SSH chain
and (b) a gapless Haldane model. Here $E_{\rm GS}$ denotes the ground-state energy.  Solid curves show DK cooling, while dashed lines mark
the Gibbs state values at $\beta_{\mathrm{th}}=4,6,10$. After one cycle, the energy density is already comparable to its target Gibbs value. We use $r=16$, $\tau=r\beta_{\mathrm{th}}/\pi$.
Other parameters are the same as in
Fig.~\ref{fig:filter_error_scaling}(d) for (a) and
Fig.~\ref{fig:twodim_cooling_mixing}(c) for (b).}
\label{fig:Gibbs_state_preparation}

\end{figure}

\section{Gibbs-state preparation at strong coupling}
So far, we have focused on ground-state preparation. Here we show the same framework at strong coupling can also be employed to prepare Gibbs states, provided that the cooling and heating probabilities satisfy a thermal balance condition. Specifically, the steady-state occupation in Eq.~\eqref{Eq:steadystaterate} follows the Fermi--Dirac distribution  $p_E^\infty=1/(e^{\beta_{\mathrm{th}}E}+1)$ if the transfer probabilities satisfy
\begin{equation}
\frac{P_\tau(-E)}{P_\tau(E)}
=
e^{-\beta_{\mathrm{th}}E}.
\label{eq:gibbs_balance_condition}
\end{equation}

This relation can be approximately satisfied by the Demkov--Kunike (DK) protocol~\cite{demkov1969hypergeometric}. Using its analytical solution in Ref.~\cite{ZlatanovVitanov2017}, we obtain (see Appendix~\ref{app:dk_thermal})
\begin{equation}
p_{E,\mathrm{DK}}^\infty
=
\frac{1}{e^{\beta_{\mathrm{th}}E}+1}
+
O\!\left(e^{-\beta_{\mathrm{th}}\Delta_0}\right),
\end{equation}
where $\beta_{\mathrm{th}}=\pi t_r$ is set by the pulse width $t_r$.

For a finite cooling cycle of period $\tau$, we use controls
\begin{align}\label{eq:DK_protocol}
f_\tau(t)
&\propto
\operatorname{sech}\!\left(\frac{t-\tau/2}{t_r}\right),
\nonumber\\
\Delta_{\mathrm A}(t)
&\propto
\tanh(r/2)
-
\tanh\!\left(\frac{t-\tau/2}{t_r}\right),
\end{align}
where $r=\tau/t_r$. The normalized controls satisfy the optimal matching condition in Eq.~\eqref{eq:optimal_matching_condition}. At fixed $r$, the cycle period therefore sets the
target inverse temperature,
\begin{equation}
\beta_{\mathrm{th}}
=
\frac{\pi\tau}{r}.
\end{equation}

Fig.~\ref{fig:Gibbs_state_preparation} compares the finite-period DK cooling dynamics with the Gibbs-state targets at $\beta_{\mathrm{th}}=4,6,10$ in (a) a gapped SSH chain
and (b) a gapless Haldane model. After one or two cycles, the energy density is already close to the corresponding Gibbs value. 
The single-cycle energy error initially decreases approximately exponentially with $\beta_{\mathrm{th}}$; increasing $r$ suppresses boundary effects and extends this regime (see Fig.~\ref{fig:appendix_gibbs_energy_error} in Appendix~\ref{app:dk_thermal}).

\section{Discussion and Conclusion}

We have developed protocols for ground- and thermal-state preparation using strongly coupled dynamical ancillas. For a class of free-fermion systems, we reduce the dynamics exactly to independent two-level scattering problems.
This reduction allows us to formulate ground-state control design as minimizing the total transfer error.  Motivated by the Landau--Zener approximation, we derive an optimal matching condition between the coupling function and the ancilla ramp in \cref{eq:optimal_matching_condition}, reducing the control design to a single function. The resulting M\"obius--Ingham family control function in \cref{Eq:mobiusprotocol,Eq:mobiusprotocol2} combines optimal gapless error scaling with an almost-exponential bound on the leading gapped heating error. Numerical benchmarks in one- and two-dimensional free-fermion systems and interacting models demonstrate the effectiveness of these controls beyond the analytically tractable setting. In the strong-coupling regimes studied, one or two cycles already achieve cooling errors comparable to their steady-state values.

The same framework also provides a route to Gibbs-state preparation. We identify a detailed balance condition in \eqref{eq:gibbs_balance_condition} on the exact transfer probabilities and show that it can be approximately satisfied by the Demkov--Kunike protocol in \eqref{eq:DK_protocol}, with the target temperature set by the characteristic width of the coupling profile.

Together, these results show that strong coupling can accelerate ground- and thermal-state preparation. The key ingredient is the joint design of the coupling function and the ancilla ramp.

A natural next step is to jointly optimize the coupling strength scaling, ancilla initial bandwidth $\Delta_0$, and control functions together. Increasing the coupling strength can reduce the Landau--Zener contribution to the passband error, but also changes the nonadiabatic heating. Similarly, increasing $\Delta_0$ improves the coverage of the system spectrum, while reducing $\Gamma_0\propto g_0^2/\Delta_0$ at fixed coupling strength. Determining how to balance these effects could further reduce the preparation time at a given target error. 

We note that finding the optimal $\Delta_A(t)$ and $f_{\tau}(t)$ choices to prepare a given profile $P_{\tau}(E)$ (in our case, a step function) is closely linked to the quantum signal processing(QSP) problem \cite{martyn2021grand, low2016methodology, low2017optimal} of finding a given sequence of discrete gate rotations to reconstruct a desired target polynomial filter. Algorithms have been constructed to find good sequences of angles efficiently \cite{haah2019product, chao2020finding}. It would be interesting to combine QSP-based control design with our approach, particularly for digital quantum platforms. Whether the resulting protocols
remain effective in interacting systems is an open question.


For interacting systems, the system--ancilla coupling operator determines which many-body states are coupled and the strengths of the corresponding transitions. Our Fermi--Hubbard benchmark, for example, uses a coupling that transfers energy to the ancillas while preserving the system particle number. This suggests optimizing not only the coupling function and ancilla ramp, but also the coupling operator itself. How to combine these choices to achieve faster cooling while respecting the required conservation laws remains an open question.

Our setup can in principle be implemented on different quantum experimental platforms, including superconducting circuits~\cite{2024Sci...383.1332M,aamir2025thermally,farrell2026preparingthermalstatesfrustrated}, trapped ions~\cite{PhysRevX.14.041046,barreiro2011open,seki2026dissipative,li2025programmable},
and neutral atoms~\cite{muniz2025repeated}. 
Notably,  cooling within one or two cycles greatly reduces both the number of ancilla resets and the associated errors.  However, decoherence and control errors remain during each cycle, as characterized in recent quantum processor benchmarks~\cite{Ransford2026Helios,Google2025BelowThreshold}. Longer cycles can suppress the intrinsic cooling error but also allow more noise to accumulate~\cite{molpeceres2026benchmarkquantumalgorithmsground}.  Imperfect ancilla initialization and noisy evolution may also shift the steady state away from the target. Suppressing intrinsic preparation errors may require additional resources, e.g., longer evolution times or more quantum gates. These can increase decoherence and accumulated gate errors. Optimizing the coupling function, ancilla ramp, and cycle period should therefore balance intrinsic preparation errors against hardware noise.

\begin{acknowledgments}
We used OpenAI's ChatGPT (GPT-4, GPT-5, GPT-6, and o-series models) to assist with editing the manuscript, developing proof strategies, and checking numerical code. The authors take full responsibility for the content of the manuscript.
Biao Lian acknowledges support by the National Science Foundation under award DMR-2141966, and National Science Foundation through the Princeton University Materials Research Science and Engineering Center DMR-2614599. 
Dmitry A. Abanin acknowledges support through a Brown Investigator Award and grant 864597 from the European Research Council (ERC), funded by the European Union's Horizon 2020 research and innovation programme.

\end{acknowledgments}

\textit{Note added.---}
While preparing this manuscript, we became aware of a related work by Yu, Chen, and Lin~\cite{2026arXiv260940328Y} on ground-state preparation in weakly interacting gapped fermionic systems without bath resets. Our optimization principles may also be applied to control design in their setting.

\bibliography{reference}

\onecolumngrid
\appendix

\section{Exact cooling dynamics and many-body fidelity}
\label{app:quadratic_reduction}

We first establish the exact two-level reduction and then derive the corresponding rate equation and an ideal single-cycle condition.

\subsection{Exact two-level reduction}
\label{app:two_level_factorization}

We consider a class of free-fermion systems with chiral symmetry, whose single-particle Hamiltonian takes the form $H_{\mathrm s}=\begin{pmatrix}0&T\\T^\dagger&0\end{pmatrix}$, where $T$ describes the hopping between the two sublattices $A$ and $B$. Let $u_n$ and $v_n$ be the normalized left and right singular vectors of $T$, satisfying $Tv_n=E_nu_n$ and $T^\dagger u_n=E_nv_n$, with $E_n>0$. Due to chiral symmetry, the single-particle eigenstates appear in pairs with energies $\pm E_n$ and eigenvectors $(u_n,\pm v_n)^T/\sqrt{2}$. The corresponding Hamiltonian can therefore be diagonalized as
\begin{equation}
\hat H_{\mathrm s}
=
\sum_{n=1}E_n
\left(
\hat c_{n,+}^\dagger\hat c_{n,+}
-
\hat c_{n,-}^\dagger\hat c_{n,-}
\right),
\end{equation}
where
\begin{equation}
\hat c_{n,\pm}^\dagger
=
\frac{1}{\sqrt2}\sum_j
\left[
u_n(j)\hat c_{A,j}^\dagger
\pm v_n(j)\hat c_{B,j}^\dagger
\right].
\label{eq:app_chiral_pair_operator}
\end{equation}

We denote the lower and upper edges of the positive-energy spectrum by $E_g$ and $E_{\max}$, respectively, and its width by $W=E_{\max}-E_g$.

For each unit cell, we introduce two fermionic ancilla modes, a sink mode $\hat s_j$ and a source mode $\hat r_j$, with energies $+\Delta_{\rm A}(t)$ and $-\Delta_{\rm A}(t)$, respectively. Both modes are locally coupled to the $A$ sublattice of the system. The ancilla Hamiltonian and the system--ancilla coupling are given by
\begin{align}
\hat H_{\mathrm a}(t)
&=
\Delta_{\mathrm A}(t)\sum_j
\left(
\hat s_j^\dagger\hat s_j
-
\hat r_j^\dagger\hat r_j
\right),
\nonumber\\
\hat H_{\mathrm c}(t)
&=
g_0f_\tau(t)\sum_j
\left[
\hat c_{A,j}^\dagger
(\hat s_j+\hat r_j)
+\mathrm{H.c.}
\right].
\label{eq:app_A_coupling}
\end{align}
Transforming both ancillas with the same singular vectors,
\begin{equation}
\hat s_n=\sum_j u_n^*(j)\hat s_j,\qquad
\hat r_n=\sum_j u_n^*(j)\hat r_j,
\label{eq:app_ancilla_bright_modes}
\end{equation}
leaves $\hat H_{\mathrm a}$ diagonal in $n$. Using $\hat c_{A,j}
=\frac{1}{\sqrt{2}}\sum_nu_n(j)\left(\hat c_{n,+}+\hat c_{n,-}\right)$ and the orthogonality of the $u_n$, the sink coupling becomes
\begin{align}
\sum_j
\hat c_{A,j}^\dagger\hat s_j
&=
\frac{1}{\sqrt{2}}
\sum_{j,n,m}
u_n^*(j)u_m(j)
\left(
\hat c_{n,+}^\dagger+\hat c_{n,-}^\dagger
\right)
\hat s_m
\nonumber\\
&=
\frac{1}{\sqrt{2}}
\sum_n
\left(
\hat c_{n,+}^\dagger+\hat c_{n,-}^\dagger
\right)
\hat s_n .
\label{eq:app_sink_diagonalization}
\end{align}
Similarly,
\begin{equation}
\sum_j
\hat c_{A,j}^\dagger\hat r_j
=
\frac{1}{\sqrt{2}}
\sum_n
\left(
\hat c_{n,+}^\dagger+\hat c_{n,-}^\dagger
\right)
\hat r_n .
\label{eq:app_source_diagonalization}
\end{equation}

Therefore,
\begin{equation}
\hat H_{\mathrm c}(t)
=
\frac{g_0f_\tau(t)}{\sqrt{2}}
\sum_n
\left[
\left(
\hat c_{n,+}^\dagger+\hat c_{n,-}^\dagger
\right)
\left(
\hat s_n+\hat r_n
\right)
+
\mathrm{H.c.}
\right].
\label{eq:app_interaction_pair_basis}
\end{equation}

Hence the full single-particle Hamiltonian can be decomposed into direct sum of different chiral-pair $n$
\begin{equation}
H_{\rm t}(t)=H_s+H_a(t)+H_{\rm c}(t)
=
\bigoplus_n
H_{E_n}(t).
\end{equation}
For a fixed positive energy $E$, we use the following basis
\begin{equation}
\left\{
|+E\rangle,\,
|s_E\rangle,\,
|r_E\rangle,\,
|-E\rangle
\right\},
\label{eq:app_four_level_basis}
\end{equation}
where $|\pm E\rangle\equiv\hat c_{\pm E}^{\dagger}|\mathrm{vac}\rangle$, $|s_E\rangle\equiv\hat s_E^{\dagger}|\mathrm{vac}\rangle$, and $|r_E\rangle\equiv\hat r_E^{\dagger}|\mathrm{vac}\rangle$, with $|\mathrm{vac}\rangle$ the fermionic vacuum.
The corresponding four-level block is
\begin{equation}
H_E(t)
=
\begin{pmatrix}
E&
G(t)&
G(t)&
0
\\
G(t)&
\Delta_{\mathrm A}(t)&
0&
G(t)
\\
G(t)&
0&
-\Delta_{\mathrm A}(t)&
G(t)
\\
0&
G(t)&
G(t)&
-E
\end{pmatrix},
\label{eq:app_four_level_H}
\end{equation}
where $G(t)=\frac{g_0f_{\tau}(t)}{\sqrt{2}}$. This matrix can be further decomposed as
\begin{equation}
H_E(t)
=
H_E^+(t)\otimes I
+
I\otimes H_E^-(t),
\label{eq:app_kronecker_sum}
\end{equation}
where each of the $H_E^\pm(t)$ takes our familiar form
\begin{equation}
H_E^+(t)
=
\begin{pmatrix}
\dfrac{E+\Delta_{\mathrm A}(t)}{2}
&
G(t)
\\ 
G(t)
&
-\dfrac{E+\Delta_{\mathrm A}(t)}{2}
\end{pmatrix},
\label{eq:app_HEplus}
\end{equation}
and
\begin{equation}
H_E^-(t)
=
\begin{pmatrix}
\dfrac{E-\Delta_{\mathrm A}(t)}{2}
&
G(t)
\\ 
G(t)
&
-\dfrac{E-\Delta_{\mathrm A}(t)}{2}
\end{pmatrix}.
\label{eq:app_HEminusappendix}
\end{equation}

Notice that $I\otimes H_E^-(t)$ couples $|+E\rangle\leftrightarrow|s_E\rangle$ and $|-E\rangle\leftrightarrow|r_E\rangle$, which are both swapping cooling channel while $H_E^+(t)\otimes I$ couples $|+E\rangle\leftrightarrow |r_E\rangle$ and $|-E\rangle\leftrightarrow|s_E\rangle$, which are non-resonant heating channel.

Therefore, the complete propagator factorizes exactly,
\begin{equation}
U_E(\tau)
=
S^+(E)\otimes S^-(E),
\label{app:app_propagator_factorization}
\end{equation}
where
\begin{equation}
S^\pm(E)
=
\mathcal T
\exp
\left[
-i
\int_0^\tau
H_E^\pm(t)\,dt
\right]=
\begin{pmatrix}
\alpha_\pm(E) & \beta_\pm(E)
\\
-\beta_\pm^*(E) & \alpha_\pm^*(E)
\end{pmatrix},
\label{eq:app_Spm}
\end{equation}
Notice that  $H_E^+(t)=\sigma_x H_{-E}^-(t)\sigma_x$,  we have $|\beta_+(E)|^2=|\beta_-(-E)|^2$.
We denote the transfer probability by $P_\tau(E)=|\beta_-(E)|^2$, so that $P_\tau(-E)=|\beta_+(E)|^2$.

Thus, the complete four-mode dynamics of each chiral pair is reduced exactly to two independent two-level scattering problems.

\subsection{Rate Equation and Single-cycle condition}
\label{app:pair_rate_equation}

At the beginning of every cycle, the sink is reset to the empty state and the source to the occupied state,
\begin{equation}\label{Eq:ancilla_initial}
\langle \hat s_E^\dagger\hat s_E\rangle=0,
\qquad
\langle \hat r_E^\dagger\hat r_E\rangle=1.
\end{equation}
The initial ancillas are in a product state, uncorrelated with the system and with each other; in particular, $\langle\hat s_E^\dagger\hat r_E\rangle=0$.

Let $\hat n_{\pm E}=\hat c_{\pm E}^\dagger\hat c_{\pm E},n_{\pm E} =\langle \hat n_{\pm E}\rangle$, and allow a general coherence between the two system modes $\chi_E
=\left\langle\hat c_{+E}^\dagger\hat c_{-E}\right\rangle.$ Expanding the Kronecker product $U_E=S^+\otimes S^-$gives
\begin{equation}
U_E
=
\begin{pmatrix}
\alpha_+\alpha_- &
\alpha_+\beta_- &
\beta_+\alpha_- &
\beta_+\beta_-
\\
-\alpha_+\beta_-^* &
\alpha_+\alpha_-^* &
-\beta_+\beta_-^* &
\beta_+\alpha_-^*
\\
-\beta_+^*\alpha_- &
-\beta_+^*\beta_- &
\alpha_+^*\alpha_- &
\alpha_+^*\beta_-
\\
\beta_+^*\beta_-^* &
-\beta_+^*\alpha_-^* &
-\alpha_+^*\beta_-^* &
\alpha_+^*\alpha_-^*
\end{pmatrix}.
\label{eq:full_tensor_propagator}
\end{equation}
The annihilation operators therefore evolve as
\begin{equation}
\begin{pmatrix}
\hat c_{+E}'\\
\hat s_E'\\
\hat r_E'\\
\hat c_{-E}'
\end{pmatrix}
=
U_E
\begin{pmatrix}
\hat c_{+E}\\
\hat s_E\\
\hat r_E\\
\hat c_{-E}
\end{pmatrix}.
\end{equation}

The positive-energy annihilation operator evolves as
\begin{equation}
\hat c_{+E}'
=
\alpha_+\alpha_-\,\hat c_{+E}
+
\alpha_+\beta_-\,\hat s_E
+
\beta_+\alpha_-\,\hat r_E
+
\beta_+\beta_-\,\hat c_{-E}.
\label{eq:dplus_evolution}
\end{equation}
Hence, one has
\begin{align}
n_{+E}'&=
|\alpha_+\alpha_-|^2n_{+E}
+|\beta_+\alpha_-|^2
+|\beta_+\beta_-|^2n_{-E}
+2\operatorname{Re}\!\left[
(\alpha_+\alpha_-)^*\beta_+\beta_-\chi_E\right],
\label{eq:nplus_general}\\
n_{-E}'&=
|\beta_+\beta_-|^2n_{+E}
+|\alpha_+\beta_-|^2
+|\alpha_+\alpha_-|^2n_{-E}
+2\operatorname{Re}\!\left[
(\alpha_+\alpha_-)^*\beta_+\beta_-\chi_E\right].
\label{eq:nminus_general}
\end{align}

Defining the particle-hole excitation density $p_E=(n_{+E}+1-n_{-E})/2$, the coherence contributions cancel exactly, yielding the closed rate equation
\begin{align*}
p_E^{(m+1)}
&=[1-P_\tau(E)]p_E^{(m)}+P_\tau(-E)[1-p_E^{(m)}]\\
&=\lambda_Ep_E^{(m)}+P_\tau(-E),
\qquad \lambda_E=1-P_\tau(E)-P_\tau(-E).
\end{align*}

In the first line of the rate equation, the first term describes excitations that remain after incomplete transfer, while the second describes excitations generated by heating.

For $|\lambda_E|<1$, the excitation density converges to the stationary value
\begin{equation*}
p_E^\infty=\frac{P_\tau(-E)}{P_\tau(E)+P_\tau(-E)}
=\frac{P_\tau(-E)}{1-\lambda_E}.
\end{equation*}

with
\begin{equation}
p_E^{(m)}-p_E^\infty
=
\lambda_E^m(p_E^{(0)}-p_E^\infty).
\end{equation}

When $\lambda_E=0$, we have $p_E^{(1)}=p_E^\infty=P_\tau(-E)$ independently of the initial excitation density. This is the ideal single-cycle condition. 

Notably, this exact equality is not required for useful single-cycle cooling. For $|\lambda_E|<1$,
\begin{equation*}
|p_E^{(1)}-p_E^\infty|
=|\lambda_E|\,|p_E^{(0)}-p_E^\infty|
\leq|\lambda_E|.
\end{equation*}
When $|\lambda_E|\lesssim p_E^\infty\ll1$, one cycle therefore gives an excitation density at most of the same order as its steady-state value.

The ideal cooling requires $P_\tau(E)=\Theta(E)$ over the system spectrum $\mathcal S=[-E_{\max},-E_g]\cup[E_g,E_{\max}]$, where $E_g=0$ for gapless spectra and $E_g>0$ for gapped spectra.

We quantify deviations from the ideal cooling as
\begin{equation}
\mathcal A(\tau;\mathcal S)
=
\int_{\mathcal S}
|P_\tau(E)-\Theta(E)|\,dE
=
\mathcal A_{\rm pass}+\mathcal A_{\rm stop},
\label{eq:appendixfilter_error}
\end{equation}
where
\begin{align}
\mathcal A_{\rm stop}(\tau;\mathcal S)
&=
\int_{E_g}^{E_{\max}}P_\tau(-E)\,dE,
\nonumber\\
\mathcal A_{\rm pass}(\tau;\mathcal S)
&=
\int_{E_g}^{E_{\max}}[1-P_\tau(E)]\,dE.
\label{eq:app_pass_stop_definitions}
\end{align}

\subsection{Many-body ground-state fidelity}\label{App:manybodyfidelity}

We relate the pair excitation density to the many-body ground-state fidelity.
Consider an initial Gaussian state that factorizes over the system energy eigenmodes,
\begin{equation}
\hat\rho_{\rm s}^{(0)}
=
\bigotimes_E
\left[
n_E^{(0)}|1_E\rangle\langle1_E|
+
(1-n_E^{(0)})|0_E\rangle\langle0_E|
\right],
\label{eq:app_manybody_density_matrix}
\end{equation}
with $n_{+E}^{(0)}+n_{-E}^{(0)}=1$ for each chiral pair.

The cooling channel preserves this form. The propagator in Eq.~\eqref{eq:full_tensor_propagator} gives
\begin{align}
n_{+E}'+n_{-E}'-1
&=
\left(
|\alpha_+\alpha_-|^2
+
|\beta_+\beta_-|^2
\right)
(n_{+E}+n_{-E}-1),
\nonumber\\
\chi_E'
&=
(\alpha_+\alpha_-\beta_+\beta_-)^*
(n_{+E}+n_{-E}-1).
\end{align}
Thus, $n_{+E}+n_{-E}=1$ and $\chi_E=0$ are preserved after each cycle. The system state therefore remains a product over its energy eigenmodes.

The half-filled ground state has all negative-energy modes occupied and all positive-energy modes empty.
Its fidelity after $m$ cycles is
\begin{equation}
\mathcal F_m
=
\prod_{E>0}
(1-n_{+E}^{(m)})n_{-E}^{(m)}
=
\prod_{E>0}(1-p_E^{(m)})^2.
\label{eq:app_manybody_fidelity_product}
\end{equation}
Taking the logarithm gives
\begin{equation}
-\frac{\ln\mathcal F_m}{L}
=
-\frac{2}{L}
\sum_{E>0}\ln(1-p_E^{(m)}).
\label{eq:app_log_fidelity_density}
\end{equation}
When $p_E^{(m)}\ll1$ for every mode,
\begin{equation}
-\frac{\ln\mathcal F_m}{L}
\simeq
\frac{2}{L}\sum_{E>0}p_E^{(m)}.
\end{equation}

At the steady state, $p_E^\infty=P_\tau(-E)/(1-\lambda_E)$. For $|\lambda_E|\ll1$ and $p_E^\infty\ll1$ throughout the spectrum, we obtain
\begin{equation}
-\frac{\ln\mathcal F_{\rm ss}}{L}
\simeq
\frac{2}{L}\sum_{E>0}P_\tau(-E)
\simeq
\frac{\mathcal A_{\rm stop}(\tau;\mathcal S)}{W}.
\label{eq:app_ss_fidelity_stopband}
\end{equation}
In this regime, the steady-state log-fidelity density is therefore determined by the stopband error.

The total transfer error also bounds the infidelity. Since $0\leq p_E^{(m-1)}\leq1$, the rate equation gives
\begin{equation}
p_E^{(m)}
\leq
1-P_\tau(E)+P_\tau(-E),
\qquad m\geq1.
\label{eq:app_mode_filter_error_bound}
\end{equation}
Using $1-\prod_j(1-x_j)\leq\sum_jx_j$ for $0\leq x_j\leq1$, we obtain
\begin{align}
1-\mathcal F_m
&\leq
2\sum_{E>0}p_E^{(m)}
\nonumber\\
&\leq
2\sum_{E>0}
\left[1-P_\tau(E)+P_\tau(-E)\right].
\label{eq:fidelity_excitation_bound}
\end{align}
Hence, we have
\begin{equation}
1-\mathcal F_m
\lesssim
\frac{L}{W}
\left[
\mathcal A_{\rm pass}(\tau;\mathcal S)
+
\mathcal A_{\rm stop}(\tau;\mathcal S)
\right]
=
\frac{L}{W}\mathcal A(\tau;\mathcal S).
\label{eq:app_finite_cycle_infidelity_bound}
\end{equation}
These relations motivate using the total transfer error as a scalar surrogate for the bi-objective cooling optimization.

\subsection{Translation-invariant systems}
\label{app:translationinvariant}

A similar two-level reduction is possible for translation-invariant free-fermion systems. Consider a number-conserving system with one orbital per unit cell and periodic boundaries,
\begin{equation}
\hat H_{\mathrm s}
=
\sum_{i,j}T_{i-j}\hat c_i^\dagger\hat c_j
=
\sum_k E_k\hat c_k^\dagger\hat c_k.
\end{equation}
We couple each system site locally to a sink and a source in separate stages. During the sink stage, the ancillas are initially empty, and
\begin{align}
\hat H_{\mathrm a}(t)
&=
\Delta_{\mathrm A}(t)\sum_j\hat s_j^\dagger\hat s_j,
\\
\hat H_{\mathrm c}(t)
&=
g_0f_\tau(t)\sum_j
(\hat c_j^\dagger\hat s_j+\mathrm{h.c.}).
\end{align}
Applying the Fourier transformation to the system and ancillas gives $\hat H_{\mathrm c}(t)
=g_0f_\tau(t)\sum_k
(\hat c_k^\dagger\hat s_k+\mathrm{H.c.})$. The dynamics therefore separates into independent $(\hat c_k,\hat s_k)$ sectors, with single-particle Hamiltonian
\begin{equation}
\begin{pmatrix}
E_k & g_0f_\tau(t)\\
g_0f_\tau(t) & \Delta_{\mathrm A}(t)
\end{pmatrix}
=
\frac{E_k+\Delta_{\mathrm A}(t)}{2}I
+
H_{E_k}^-(t).
\end{equation}
The identity term contributes only a common phase within each sector.

During the source stage, the system instead couples to initially occupied modes $\hat r_j$ at energy
$-\Delta_{\mathrm A}(t)$. The same argument gives $H_{E_k}^+(t)$. Thus, both stages reduce exactly to two-level scattering problems,
\begin{equation}
H_{E_k}^{\pm}(t)
=
\frac{E_k\pm\Delta_{\mathrm A}(t)}{2}\sigma_z
+
g_0f_\tau(t)\sigma_x,
\end{equation}
with coupling matrix element $V=1$.

\begin{figure}[t]
\centering
\includegraphics[height=\paperfigureheight]{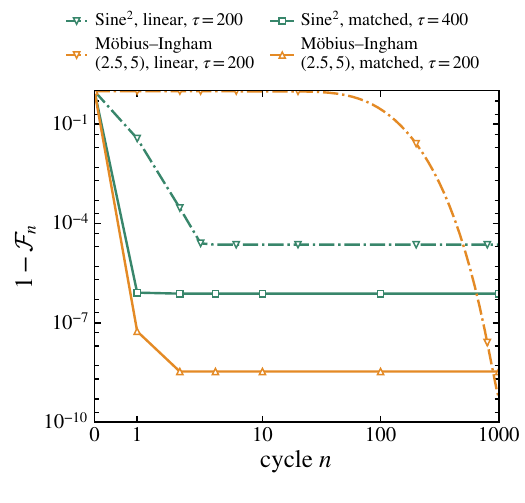}
\caption{Ground-state infidelity $1-\mathcal F_n$ versus cycle $n$ for linear (dash-dotted) and matched (solid) ancilla ramps in a gapped SSH chain with $L=100$.  Other parameters are the same as in Fig.~\ref{fig:filter_error_scaling}(d).}
\label{fig:linear_matched_ramps}
\end{figure}

\section{Optimal matching condition between the coupling and the ancilla ramp}
\label{app:optimal_matching}

For a given coupling function, we choose the ancilla ramp to suppress the passband error in the cooling sector $H_E^-(t)$. The local Landau--Zener approximation provides a design principle for this matching.

\begin{equation}
H_E^-(t)
=
\begin{pmatrix}
\dfrac{E-\Delta_{\mathrm A}(t)}{2}
&
G(t)
\\ 
G(t)
&
-\dfrac{E-\Delta_{\mathrm A}(t)}{2}
\end{pmatrix}.
\label{eq:app_HEminus}
\end{equation}
where  $G(t)= g_0 f_{\tau}(t)V$. Without loss of generality, we take $V$ to be real and positive, since its phase can be absorbed into the definition of one of the two basis states.

For $0<E<\Delta_0$, the resonance time is defined by $\Delta_{\mathrm A}(t_E)=E$.
Linearizing the ramp about $t_E$ and replacing the coupling by its value at the crossing gives the local Landau--Zener Hamiltonian~\cite{zener1932non},
\begin{equation}
H_E^-(t)\simeq H_{\mathrm{LZ}}(t)
=
\begin{pmatrix}
v(t-t_E)/2 & G\\
G & -v(t-t_E)/2
\end{pmatrix},
\end{equation}
where $v=-\dot{\Delta}_{\mathrm A}(t_E)>0$ and $G=g_0Vf_\tau(t_E)$. The corresponding Landau--Zener estimate of the passband error is
\begin{equation}
1-P_\tau(E)\simeq e^{-\Gamma_E},
\qquad
\Gamma_E=
\frac{2\pi g_0^2|V|^2|f_\tau(t_E)|^2}
{-\dot{\Delta}_{\mathrm A}(t_E)}.
\end{equation}

Thus, within the local Landau--Zener approximation, cooling optimization is equivalent to maximizing the smallest Landau--Zener exponent $\Gamma_E$. Since the prefactor $2\pi g_0^2V^2$ is independent of $E$, the optimization reduces to
\begin{equation}
\max_{\Delta_{\mathrm A}(t)}
\;
\min_{0<t<\tau}
\frac{
|f(t)|^2
}{
-\dot{\Delta}_{\mathrm A}(t)
},
\label{eq:minimax_LZ_problem}
\end{equation}
subject to $\dot{\Delta}_{\mathrm A}(t)\leq0 $, $\Delta_A(0)=\Delta_0$, $\quad \Delta_A(\tau)=0$ and $\int_0^\tau |f_\tau(t)|^2dt=1$.

The minimum of this ratio cannot exceed its average weighted by the ramping speed $-\dot{\Delta}_{\mathrm A}(t)$, giving~\cite{1996JMagR.120..133T,GARWOOD2001155,optimalrelation2002}
\begin{align}
    \min_{0<t<\tau}\frac{|f_\tau(t)|^2}{-\dot{\Delta}_{\mathrm A}(t)}
&\leq\frac{\int_0^\tau |f_\tau(t)|^2\,dt}{\int_0^\tau [-\dot{\Delta}_{\mathrm A}(t)]\,dt}
=\frac{1}{\Delta_0}.
\end{align}
The bound is saturated if and only if the local Landau--Zener exponent is uniform throughout the ramp $\frac{|f_\tau(t)|^2}
{-\dot{\Delta}_{\mathrm A}(t)}=\frac{1}{\Delta_0}$, which yields the optimal matching condition
\begin{equation}
-\dot{\Delta}_{\mathrm A}^{\mathrm{opt}}(t)
=
\Delta_0 |f_\tau(t)|^2 .
\label{eq:general_optimal_velocity}
\end{equation}

The corresponding Landau--Zener exponent is independent of $E$,
\begin{equation*}
\Gamma_E=\Gamma_0,
\qquad
\Gamma_0=\frac{2\pi g_0^2|V|^2}{\Delta_0}.
\end{equation*}
Within the same approximation, the integrated passband error is
\begin{equation*}
\mathcal A_{\rm pass}(\tau;\mathcal S)
\simeq
\int_{E_g}^{E_{\max}}e^{-\Gamma_0}\,dE
=We^{-\Gamma_0}.
\end{equation*}
Corrections to the Landau--Zener estimate at finite cycle period are discussed in
Appendix~\ref{App:singlecycleerror}. With the ramp determined by $f_\tau(t)$, the remaining design problem is to suppress the stopband error through this function.

Fig.~\ref{fig:linear_matched_ramps} compares cooling with linear $\Delta_A(t)=\Delta_0(1-t/\tau)$ and optimally matched ancilla ramps for sine-squared and M\"obius--Ingham coupling functions. For both coupling functions, the matched ramps achieve low infidelity close to their steady-state value after a single cycle, whereas the linear ramps require many more cycles and a longer total physical time $n_{\rm mix}\tau$ to approach their steady state.

\begin{figure*}[t]
\centering
\includegraphics[height=\paperfigureheight]{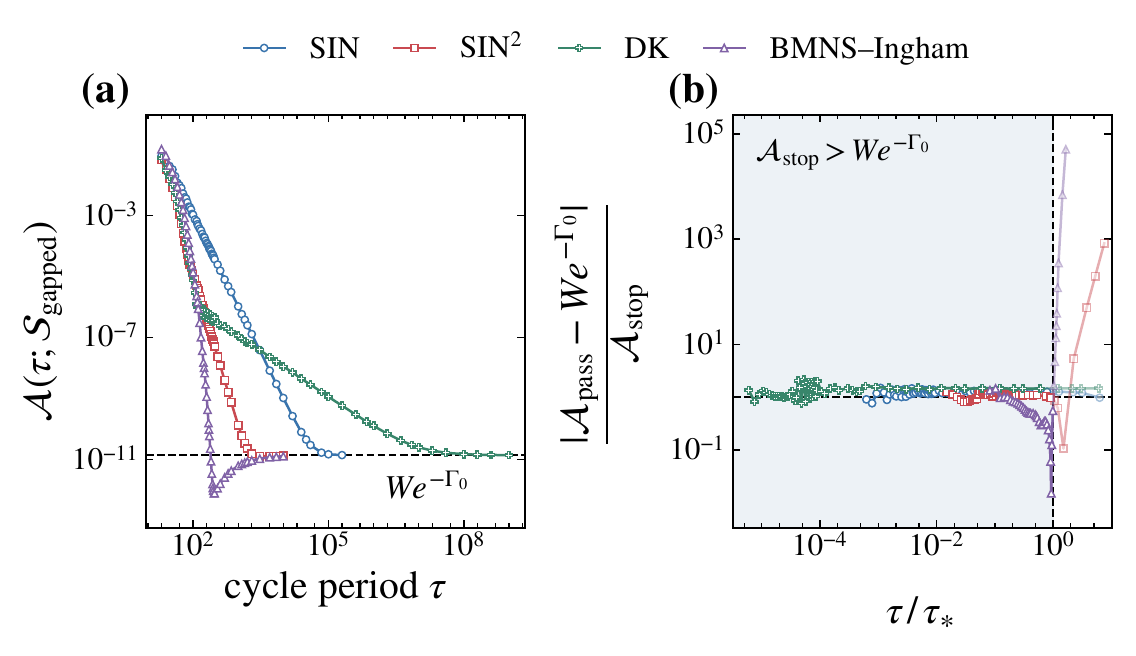}
\caption{Total transfer error and passband correction for the sine (SIN) and sine-squared (SIN$^2$) protocols defined in Appendix~\ref{app:benchmark_protocols}, the Demkov--Kunike(DK) protocol~\cite{demkov1969hypergeometric}, and the BMNS--Ingham protocol. All controls satisfy the optimal matching condition.
(a) $\mathcal A=\mathcal A_{\rm pass}+\mathcal A_{\rm stop}$ versus cycle
period $\tau$. The horizontal dashed line marks the Landau--Zener
estimate $We^{-\Gamma_0}$.
(b) $|\mathcal A_{\rm pass}-We^{-\Gamma_0}|/\mathcal A_{\rm stop}$ versus
$\tau/\tau_\ast$, where $\mathcal A_{\rm stop}(\tau_\ast)=We^{-\Gamma_0}$.
The shaded region has $\tau<\tau_\ast$, the vertical dashed line marks
$\tau=\tau_\ast$, and the horizontal dashed line marks unity.
Panel (b) includes points with $\mathcal A_{\rm stop}\ge10^{-18}$ and
$\tau/\tau_\ast\le10$.
Parameters are $E_g=0.5$, $E_{\max}=1.5$, $\Delta_0=2.25$,
$\Gamma_0=25$, and $|V|^2=1/2$.}
\label{fig:passband_correction}
\end{figure*}

\begin{figure}[t]
\centering
\includegraphics[height=\paperfigureheight]{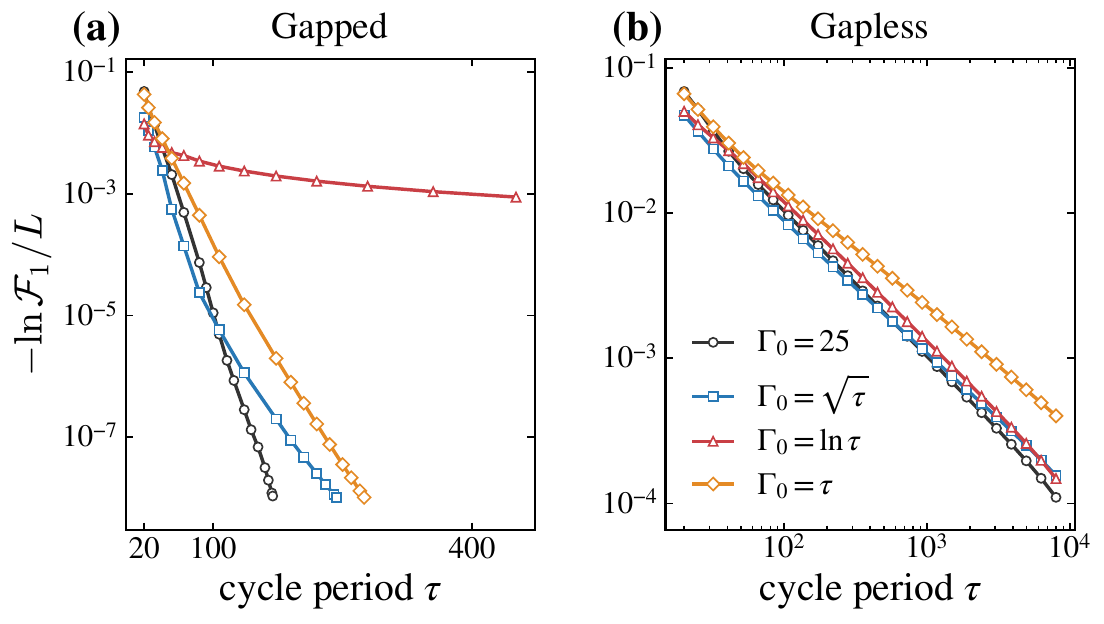}
\caption{Single-cycle log-fidelity density $-\ln\mathcal F_1/L$ for different coupling strength scaling in (a) gapped and (b) gapless SSH chains, with $(t_1^{(s)},t_2^{(s)})=(1,0.5)$ and $(1,1)$, respectively. Both chains have $L=5000$ system sites, open boundaries, and initial covariance $C_0=I/2$. We use the M\"obius--Ingham function with $(c_1,c_2)=(2.5,5)$ and compare $\Gamma_0=25$, $\sqrt{\tau}$, $\ln\tau$, and $\tau$, with $g_0(\tau)=\sqrt{\Gamma_0(\tau)\Delta_0/\pi}$. All curves use $\Delta_0=E_{\max}/(1-\alpha)$ with $\alpha=1/3$, $|V|^2=1/2$.}
\label{fig:coupling_strength_scaling}
\end{figure}

\section{Single-cycle condition error and coupling strength}
\label{App:singlecycleerror}

When the passband error is at most of the same order as the stopband error, suppressing the stopband error also controls the total transfer error and the accuracy of single-cycle cooling.
We establish this connection below and then discuss the limitations imposed by a fixed coupling strength.

\paragraph{Finite-period passband corrections.}
For controls satisfying the optimal matching condition, the Landau--Zener analysis gives the large-period estimate
\begin{equation}
\mathcal A_{\rm pass}(\tau;\mathcal S)
\simeq We^{-\Gamma_0}.
\label{eq:matched_passband}
\end{equation}

We consider the low-error window
\begin{align}\label{Eq:stopwindow}
    We^{-\Gamma_0}\lesssim\mathcal A_{\rm stop}\ll W,
\end{align}
where this estimated passband contribution is less than the stopband error.

At finite cycle periods, the total transfer error can be written exactly as
\begin{equation*}
\mathcal A
=
\mathcal A_{\rm stop}
+
We^{-\Gamma_0}
+
\left(\mathcal A_{\rm pass}-We^{-\Gamma_0}\right).
\end{equation*}
The sum of the first two terms is of order $\mathcal A_{\rm stop}$ in the window above. 

 The total error therefore remains of the same order if the finite-period correction satisfies
\begin{equation}
\left|\mathcal A_{\rm pass}-We^{-\Gamma_0}\right|
=
O(\mathcal A_{\rm stop}).
\label{eq:app_passband_correction_condition}
\end{equation}
Under this condition,
\begin{equation}
\mathcal A_{\rm stop}
\leq
\mathcal A
=
O(\mathcal A_{\rm stop}),
\label{eq:app_total_stop_comparison}
\end{equation}
so suppressing the stopband error also suppresses the total transfer error.

Fig.~\ref{fig:passband_correction} tests this condition for the sine, sine-squared, DK, and BMNS--Ingham controls introduced in Appendix~\ref{app:benchmark_protocols}. Panel (a) shows that the total transfer error decreases with the cycle period and approaches the Landau--Zener scale $We^{-\Gamma_0}$ at long periods, where the residual passband error dominates. 

Panel (b) shows $|\mathcal A_{\rm pass}-We^{-\Gamma_0}|/\mathcal A_{\rm stop}$ versus $\tau/\tau_\ast$ with $\mathcal A_{\rm stop}(\tau_\ast)=We^{-\Gamma_0}$. Within the low-error window defined in Eq.~\eqref{Eq:stopwindow}, $We^{-\Gamma_0}\lesssim\mathcal A_{\rm stop}\ll W$, this ratio remains $O(1)$ or smaller.  The results therefore support Eq.~\eqref{eq:app_total_stop_comparison}  in this regime.

The long-period saturation scale is set by the fixed choice of $\Gamma_0$ and can be lowered by increasing
the coupling strength. Increasing $\Gamma_0$ does not, however, necessarily improve cooling at shorter periods, where the stopband error dominates. We leave the joint optimization of $\Gamma_0(\tau)$ and the coupling function for future work, as discussed in the main text.

\paragraph{Single-cycle cooling.}
We consider the period range  $We^{-\Gamma_0}\lesssim\mathcal A_{\rm stop}\ll W$ in which Eq.~\eqref{eq:app_passband_correction_condition}  holds, so that $\mathcal A=O(\mathcal A_{\rm stop})$.

Using $|p_E^{(1)}-p_E^\infty|\leq|\lambda_E|$ and $|\lambda_E|\leq1-P_\tau(E)+P_\tau(-E)$, we obtain
\begin{equation}
\begin{aligned}
\int_{E_g}^{E_{\max}}
|p_E^{(1)}-p_E^\infty|\,dE
&\leq
\int_{E_g}^{E_{\max}}|\lambda_E|\,dE\\
&\leq
\mathcal A
=
O(\mathcal A_{\rm stop}).
\end{aligned}
\label{eq:single_cycle_steady_precision}
\end{equation}
Since $\int_{E_g}^{E_{\max}}p_E^{(1)}\,dE\leq\mathcal A$ from Eq.~\eqref{eq:app_mode_filter_error_bound}, one cycle therefore gives a total excitation error at most of the same order as its steady-state value within this period range.

\paragraph{Constraints on the coupling strength. }

Classical approximation of $\operatorname{sgn}(E)$ by entire functions of exponential type gives an error that is exponentially small in $E_g\tau$~\cite{EremenkoYuditskii2007}. This provides an exponential error scale for comparison. Achieving such suppression in our cooling problem, however, also requires an increasing coupling budget.

Restricting the nonlinear Parseval identity to the passband gives~\cite{Medvedevnogobound}
\begin{equation}
\int_{E_g}^{E_{\max}}-\ln[1-P_\tau(E)]\,dE
\leq
2\pi\int_0^\tau|G(t)|^2\,dt
=
\Gamma_0\Delta_0.
\label{App:Parseval_identity}
\end{equation}
Jensen's inequality then yields
\begin{equation}
\mathcal A_{\rm pass}
\geq
W\exp\!\left[-\frac{\Gamma_0\Delta_0}{W}\right].
\label{eq:app_passband_budget_bound}
\end{equation}
For a fixed gapped spectrum, suppose that $\mathcal A(\tau;\mathcal S_{\rm gapped})
\leq CW e^{-aE_g\tau}$, with $\tau$-independent $C,a>0$. Since $\mathcal A\geq\mathcal A_{\rm pass}$, this requires
\begin{equation}
2\pi g_0(\tau)^2|V|^2
=
\Gamma_0(\tau)\Delta_0
\geq
W(aE_g\tau-\ln C).
\label{eq:app_coupling_budget_scaling}
\end{equation}
Thus, $g_0(\tau)^2$ must grow at least linearly with $\tau$ for fixed $V$.

This necessary condition does not guarantee improved cooling at finite periods.
Fig.~\ref{fig:coupling_strength_scaling} illustrates this for the M\"obius--Ingham protocol by comparing $\Gamma_0=25$, $\sqrt{\tau}$, $\ln\tau$, and $\tau$. We therefore use a fixed, sufficiently large coupling strength over the precision range studied and leave the joint optimization of $\Gamma_0(\tau)$ and the coupling function for future work.

\section{Stopband error scaling}
\label{subsec:optimal_gapless_gapped}

With the ancilla ramp determined by the optimal matching condition, we now analyze how the coupling function suppresses the stopband error. 

Following the main text, we introduce a fixed cutoff $0<\Lambda<E_{\max}$ and decompose
\begin{align}
\mathcal A_{\rm stop}(\tau;\mathcal S_{\rm gapless})
&=
\int_0^\Lambda P_\tau(-E)\,dE+
\int_\Lambda^{E_{\max}}P_\tau(-E)\,dE.
\label{eq:app_gapless_stopband_decomposition}
\end{align}
The first integral represents the transition error near $E=0$. The second is the stopband error of  a gapped spectrum with $E_g=\Lambda$. When this second contribution is subleading, the crossover determines the gapless stopband scaling.

We first analyze how the coupling function suppresses this gapped contribution. We then identify the conditions for optimal gapless scaling when heating away from zero is sufficiently suppressed.

\subsection{Gapped systems}\label{app:gapped_switching}

For gapped spectra, modes with $|E|<E_g$ do not contribute to the stopband error. We therefore analyze the heating sector $H_E^+(t)$ in Eq.~\eqref{eq:app_HEplus}, whose transfer probability gives the exact stopband error,
\begin{align}
\mathcal A_{\rm stop}(\tau;\mathcal S_{\rm gapped})
&=
\int_{E_g}^{E_{\max}}P_\tau(-E)\,dE
=
\int_{E_g}^{E_{\max}}|\beta_+(E)|^2\,dE.
\label{eq:gapped_error_heating_representation}
\end{align}

In the following, we will use first-order adiabatic perturbation theory to determine how endpoint smoothness and Fourier decay control the stopband error.

As discussed above, $H_E^+(t)\otimes I$ couples $|+E\rangle\leftrightarrow|r_E\rangle$ and $|-E\rangle\leftrightarrow|s_E\rangle$ separately. It is therefore sufficient to analyze the $\{|+E\rangle,|r_E\rangle\}$ sector, where the evolving state takes the form $|\psi_E(t)\rangle =a_E(t)|+E\rangle+b_E(t)|r_E\rangle$, with initial state $|\psi_E(0)\rangle=|r_E\rangle$. We take $G(0)=G(\tau)=0$ and define the dimensionless ramp as $q(x)=\Delta_{\rm A}(\tau x)/\Delta_0$.

The Hamiltonian $H_E^+(t)$ is diagonalized by
\begin{equation}
R_E(t)
=
\begin{pmatrix}
\cos\vartheta_E(t) & -\sin\vartheta_E(t)\\
\sin\vartheta_E(t) & \cos\vartheta_E(t)
\end{pmatrix},
\end{equation}
where
\begin{equation}
\vartheta_E(t)
=
\frac{1}{2}\arctan
\left[
\frac{2G(t)}{E+\Delta_{\rm A}(t)}
\right].
\end{equation}
The corresponding eigenvalues are $\pm\Omega_E^+(t)/2$, with
\begin{equation}
\Omega_E^+(t)
=
\sqrt{
[E+\Delta_{\rm A}(t)]^2+4G(t)^2
}.
\end{equation}

Defining the accumulated phase
\begin{equation}
\Phi_E^+(t)
=
\int_0^t dt'\,\Omega_E^+(t'),
\label{eq:gapped_plus_phase}
\end{equation}
we introduce the amplitudes $c_{1,E}(t)$ and $c_{2,E}(t)$ through
\begin{equation}
\begin{pmatrix}
a_E(t)\\
b_E(t)
\end{pmatrix}
=
R_E(t)
\begin{pmatrix}
c_{1,E}(t)e^{-i\Phi_E^+(t)/2}\\
c_{2,E}(t)e^{i\Phi_E^+(t)/2}
\end{pmatrix}.
\label{eq:gapped_rotating_basis}
\end{equation}
Substituting into the Schr\"odinger equation gives
\begin{align}
\dot c_{1,E}(t)
&=
\dot\vartheta_E(t)
e^{i\Phi_E^+(t)}
c_{2,E}(t),
\nonumber\\
\dot c_{2,E}(t)
&=
-\dot\vartheta_E(t)
e^{-i\Phi_E^+(t)}
c_{1,E}(t).
\label{eq:gapped_plus_coefficient_equations}
\end{align}

Since $G(0)=G(\tau)=0$, the rotation satisfies $R_E(0)=R_E(\tau)=I$. Together with the initial state $|r_E\rangle$,
this gives $c_{1,E}(0)=0$ and $c_{2,E}(0)=1$. The heating probability at the end of the cycle is therefore
\begin{equation}
|\beta_+(E)|^2
=
|a_E(\tau)|^2
=
|c_{1,E}(\tau)|^2.
\end{equation}

At first order in adiabatic perturbation theory, we set $c_{2,E}(t)\simeq1$ in Eq.~\eqref{eq:gapped_plus_coefficient_equations}, obtaining

\begin{equation}
c_{1,E}^{(1)}(\tau)
=
\int_0^\tau
dt\,
\dot\vartheta_E(t)
e^{i\Phi_E^+(t)}.
\label{eq:gapped_first_order_transition_amplitude}
\end{equation}
Using $\frac{d}{dt}e^{i\Phi_E^+(t)}=i\Omega_E^+(t)e^{i\Phi_E^+(t)}$ and integration by parts, together with $\vartheta_E(0)=\vartheta_E(\tau)=0$, gives
\begin{equation}
c_{1,E}^{(1)}(\tau)
=
-\frac{i}{2}
\int_0^\tau
\Omega_E^+(t)
\arctan\left[
\frac{2G(t)}{E+\Delta_{\rm A}(t)}
\right]
e^{i\Phi_E^+(t)}\,dt.
\label{eq:gapped_weighted_oscillatory_integral}
\end{equation}

To express this amplitude as a Fourier integral, we introduce
\begin{equation}
u_{E,\tau}(x)
=
\int_0^x\Omega_E^+(\tau y)\,dy,
\qquad
U_{E,\tau}=u_{E,\tau}(1),
\label{eq:gapped_plus_phase_variable}
\end{equation}
so that $\Phi_E^+(\tau x)=\tau u_{E,\tau}(x)$. Since $du_{E,\tau}/dx=\Omega_E^+(\tau x)\geq E_g>0$, this change of variables is invertible.

Denoting the inverse by $x=X_{E,\tau}(u)$, we obtain
\begin{equation}
c_{1,E}^{(1)}(\tau)
=
-\frac{i\tau}{2}
\int_0^{U_{E,\tau}}
F_{E,\tau}(u)e^{i\tau u}\,du,
\label{eq:gapped_exact_oscillatory_integral}
\end{equation}
where
\begin{equation}
F_{E,\tau}(u)
=
\left.
\arctan\left[
\frac{\sqrt{2}g_0s(x)}
{\sqrt{\tau}[E+\Delta_0q(x)]}
\right]
\right|_{x=X_{E,\tau}(u)}.
\label{eq:gapped_fourier_amplitude}
\end{equation}

We now determine how the endpoint smoothness of $s(x)$ controls the decay of this Fourier integral.

Suppose the first nonvanishing endpoint derivative of $s(x)$ is of order $m\geq1$:
\begin{equation}
s^{(j)}(0)=s^{(j)}(1)=0,
\qquad j=0,\ldots,m-1,
\label{eq:gapped_boundary_order_m}
\end{equation}
with at least one of $s^{(m)}(0)$ and $s^{(m)}(1)$ nonzero.

Near either endpoint, expanding the arctangent  $\arctan z= z+O(z^3)$ gives
\begin{equation}
F_{E,\tau}(u)
=
\left.
\frac{\sqrt{2}g_0}{\sqrt{\tau}}
\frac{s(x)}{E+\Delta_0q(x)}
\right|_{x=X_{E,\tau}(u)}
+
O\left(\frac{s(x)^3}{\tau^{3/2}}\right).
\label{eq:gapped_F_endpoint_expansion}
\end{equation}

Using
\begin{equation}
\frac{\partial}{\partial u}
=
\frac{1}{\Omega_E^+(\tau x)}
\frac{\partial}{\partial x},
\label{eq:gapped_u_derivative}
\end{equation}
the endpoint conditions on $s(x)$ give
\begin{equation}
F_{E,\tau}^{(j)}(0)
=
F_{E,\tau}^{(j)}(U_{E,\tau})
=
0,
\qquad j=0,\ldots,m-1.
\label{eq:gapped_F_lower_derivatives}
\end{equation}
Here derivatives of $F_{E,\tau}$ are taken with respect to $u$.

At order $m$, only the derivative acting entirely on $s(x)$ survives. Using $q(0)=1$ and $q(1)=0$, we find
\begin{align}
F_{E,\tau}^{(m)}(0)
&=
\frac{\sqrt{2}g_0}{\sqrt{\tau}}
\frac{s^{(m)}(0)}{(E+\Delta_0)^{m+1}},
\\
F_{E,\tau}^{(m)}(U_{E,\tau})
&=
\frac{\sqrt{2}g_0}{\sqrt{\tau}}
\frac{s^{(m)}(1)}{E^{m+1}}.
\end{align}

Repeated integration by parts then gives
\begin{align}
\int_0^{U_{E,\tau}}
F_{E,\tau}(u)e^{i\tau u}\,du
= &
\frac{(-1)^m}{(i\tau)^{m+1}}
\left[
F_{E,\tau}^{(m)}(u)e^{i\tau u}
\right]_0^{U_{E,\tau}}
+
\frac{(-1)^{m+1}}{(i\tau)^{m+1}}
\int_0^{U_{E,\tau}}
F_{E,\tau}^{(m+1)}(u)e^{i\tau u}\,du.
\label{eq:gapped_first_surviving_boundary}
\end{align}
Hence, 
\begin{align}
c_{1,E}^{(1)}(\tau)
= &
-\frac{i^m g_0}{\sqrt{2}\,\tau^{m+1/2}}
\left[
\frac{s^{(m)}(1)}{E^{m+1}}e^{i\tau U_{E,\tau}}
-
\frac{s^{(m)}(0)}{(E+\Delta_0)^{m+1}}
\right]+
O\left(\tau^{-m-3/2}\right).
\label{eq:gapped_boundary_amplitude}
\end{align}
Consequently, the first-order stopband error satisfies
\begin{align}
\mathcal A_{\rm stop}^{(1)}
(\tau;\mathcal S_{\rm gapped})
&=
\int_{E_g}^{E_{\max}}
|c_{1,E}^{(1)}(\tau)|^2\,dE
\nonumber\\
&\leq
W\sup_{E\in[E_g,E_{\max}]}
|c_{1,E}^{(1)}(\tau)|^2
\nonumber\\
&=
O\left(\tau^{-(2m+1)}\right).
\label{eq:app_first_order_stopband}
\end{align}
It is straightforward to check that the same stopband-error scaling also holds for $m=0$, giving
$\mathcal A_{\rm stop}^{(1)}=O(\tau^{-1})$. The sine and sine-squared coupling functions have $m=1$ and $m=2$, respectively, giving the $\tau^{-3}$ and $\tau^{-5}$ bounds discussed in the main text.

When higher-order corrections are subleading, this bound also controls the exact stopband error.
Within the period range specified in Eq.~\eqref{eq:app_total_stop_comparison}, the total transfer error therefore satisfies
\begin{equation}
\mathcal A(\tau;\mathcal S_{\rm gapped})
=
O\left(\mathcal A_{\rm stop}
(\tau;\mathcal S_{\rm gapped})\right)
=
O\left(\tau^{-(2m+1)}\right).
\label{eq:gapped_filter_power_m}
\end{equation}

For an infinitely flat coupling function,
\begin{equation}
s^{(n)}(0)=s^{(n)}(1)=0,
\qquad n=0,1,2,\ldots,
\label{eq:gapped_infinite_flatness}
\end{equation}
all finite-order boundary terms vanish. The decay of the first-order stopband error is then
determined by the Fourier tail of $F_{E,\tau}(u)$ in Eq.~\eqref{eq:gapped_exact_oscillatory_integral}.

To understand how choosing $s(x)$ can suppress the Fourier tail of $F_{E,\tau}(u)$ in
Eq.~\eqref{eq:gapped_exact_oscillatory_integral}, we consider a large-$\tau$ expansion at fixed
$s(x)$, $q(x)$. We have
\begin{equation}
\Omega_E^+(\tau x)
=
E+\Delta_0q(x)
+
\frac{g_0^2s(x)^2}
{\tau[E+\Delta_0q(x)]}
+
O(\tau^{-2}),
\label{eq:gapped_omega_expansion}
\end{equation}
and
\begin{equation}
\arctan\left[
\frac{\sqrt{2}g_0s(x)}
{\sqrt{\tau}[E+\Delta_0q(x)]}
\right]
=
\frac{\sqrt{2}g_0s(x)}
{\sqrt{\tau}[E+\Delta_0q(x)]}
+
O(\tau^{-3/2}).
\label{eq:gapped_arctan_expansion}
\end{equation}

The dynamical phase therefore becomes
\begin{equation}
\Phi_E^+(\tau x)
=
\tau v_E(x)+\chi_E(x)+O(\tau^{-1}),
\label{eq:gapped_phase_expansion}
\end{equation}
where
\begin{align}
v_E(x)
&=
\int_0^x[E+\Delta_0q(y)]\,dy,
\nonumber\\
\chi_E(x)
&=
g_0^2\int_0^x
\frac{s(y)^2}{E+\Delta_0q(y)}\,dy.
\label{eq:gapped_fixed_phase_functions}
\end{align}
Since $v_E'(x)=E+\Delta_0q(x)\geq E_g>0$, we can use $v=v_E(x)$ as the integration variable.
Let $V_E=v_E(1)$ and denote the inverse map by $x=Y_E(v)$. Keeping the leading terms in the integrand of
Eq.~\eqref{eq:gapped_weighted_oscillatory_integral} then gives
\begin{equation}
c_{1,E}^{(1)}(\tau)
\approx
-\frac{ig_0\sqrt{\tau}}{\sqrt{2}}
\int_0^{V_E}
\mathcal W_E(v)e^{i\tau v}\,dv,
\label{eq:gapped_fixed_fourier_form}
\end{equation}
where
\begin{equation}
\mathcal W_E(v)
=
\left.
\frac{s(x)}{E+\Delta_0q(x)}
e^{i\chi_E(x)}
\right|_{x=Y_E(v)}.
\label{eq:gapped_W_definition}
\end{equation}
The function $\mathcal W_E$ is independent of $\tau$. Defining its Fourier transform by
$\widetilde{\mathcal W}_E(k)
=\int_0^{V_E}\mathcal W_E(v)e^{ikv}\,dv$, we obtain
\begin{equation}
c_{1,E}^{(1)}(\tau)
\approx
-\frac{ig_0\sqrt{\tau}}{\sqrt{2}}\,
\widetilde{\mathcal W}_E(\tau).
\label{eq:gapped_c1_W_fourier}
\end{equation}

For fixed $s(x)$, $q(x)$, the function $\mathcal W_E$ is independent of $\tau$ and has compact
support. The Paley--Wiener theorem~\cite{paley1934fourier} therefore rules out the bound $|\widetilde{\mathcal W}_E(k)|\leq Ce^{-a|k|}$ for all sufficiently large $|k|$, with fixed $C,a>0$.

As a concrete example, the BMNS--Ingham function $s_{\mathrm I}(x)$ introduced in Sec.~\ref{App:inghamfunction} combines flatness at both endpoints with almost-exponential Fourier decay. We use this function as a starting point for constructing our controls. Its derivatives satisfy~\cite{Bachmanningham}
\begin{equation}
\|s_{\mathrm I}^{(n)}\|_\infty
\leq
C_sR_s^n n![\ln(e+n)]^{2n}.
\label{eq:ingham_derivative_bound}
\end{equation}
To see why this choice also suppresses the Fourier tail of $\mathcal W_E$, recall that
\begin{equation}
\mathcal W_E(v)
=
\left.
\frac{s(x)}{E+\Delta_0q(x)}
e^{i\chi_E(x)}
\right|_{x=Y_E(v)}.
\end{equation}
The function class defined by Eq.~\eqref{eq:ingham_derivative_bound} is preserved under multiplication, integration, analytic composition, and inversion~\cite{
Compositioninultradifferentiableclasse,
ultradifferentiablefunctionclass}. Since $q'(x)=-s(x)^2$, the ramp belongs to the same class as $s(x)$. Consequently,
\begin{equation}
\|\partial_v^n\mathcal W_E\|_\infty
\leq
C_WR_W^n n![\ln(e+n)]^{2n},
\end{equation}
where the constants may depend on the coupling strength and spectral gap. The endpoint flatness of $s(x)$ is also preserved in $\mathcal W_E$. Repeated integration by parts therefore gives an almost-exponential Fourier-decay bound for $\mathcal W_E$ as well.

More generally, for any function class defined by derivative bounds, and preserved under the operations above~\cite{Compositioninultradifferentiableclasse,ultradifferentiablefunctionclass},
$\mathcal W_E$ belongs to the same class as $s(x)$, and their Fourier tails obey bounds of the same scaling form.



\subsection{Gapless systems}\label{Appendix:Gaplessscaling}

For gapless spectra, modes arbitrarily close to $E=0$ also contribute to the stopband error.
When heating away from zero is subleading, the dominant contribution comes from the
transition of $P_\tau(E)$ near $E=0$. We now examine how the width of this transition determines the error scaling and we first consider a constant density of states near zero energy.

A protocol of period $\tau$ has an energy resolution of order $\delta E_\tau\gtrsim1/\tau$. For a transition region with transfer probability change of  $O(1)$, the total transfer error cannot decay faster than $1/\tau$. In the regime $\mathcal A_{\rm pass}=O(\mathcal A_{\rm stop})$, this gives
\begin{equation}
\mathcal A_{\rm stop}(\tau;\mathcal S_{\rm gapless})
\gtrsim
\frac{1}{\tau}.
\label{eq:gapless_lower_bound}
\end{equation}

We now show how this scaling follows from a dimensionless scaling limit of the corresponding two-level Schr\"odinger equation. Using the dimensionless time $x$ defined above and introducing the scaled energy $\varepsilon=\tau E$, the Schr\"odinger equation for $H_E^-(t)$ becomes
\begin{equation}
i\partial_x\psi_E
=
\left[
\frac{1}{2}
\left(
\varepsilon
-
\tau\Delta_{\rm A}(\tau x)
\right)\sigma_z
+
\tau G(\tau x)\sigma_x
\right]\psi_E.
\label{eq:gapless_scaled_schrodinger_general}
\end{equation}
At fixed $g_0$, if the rescaled controls $\tau\Delta_{\rm A}(\tau x)$ and $\tau G(\tau x)$ approach $\tau$-independent functions, the transfer probability takes the scaling form $P_\tau(\varepsilon/\tau)\to\mathcal P(\varepsilon)$, with a $\tau$-independent function $\mathcal P$. 

Using this scaling form, the contribution from the transition region can be written as
\begin{align}
\int_0^\Lambda P_\tau(-E)\,dE
&=
\frac{1}{\tau}
\int_0^{\tau\Lambda}
P_\tau(-\varepsilon/\tau)\,d\varepsilon
\nonumber\\
&\simeq
\frac{1}{\tau}
\int_0^{\tau\Lambda}
\mathcal P(-\varepsilon)\,d\varepsilon
\nonumber\\
&\le
\frac{1}{\tau}
\int_0^\infty
\mathcal P(-\varepsilon)\,d\varepsilon.
\label{eq:gapless_scaled_area}
\end{align}
Since this integral is finite and independent of $\tau$, the transition-region contribution scales as $1/\tau$.

This argument extends to gapless systems with a nonconstant density of states. The corresponding stopband error is
\begin{equation}
\mathcal A_{\rm stop}(\tau;\mathcal S_{\rm gapless,\rho})
=
\int_0^{E_{\max}}
dE\,\rho(-E)P_\tau(-E).
\end{equation}
For a low-energy density of states $\rho(E)\simeq\rho_0|E|^\nu$ with $\nu>-1$, the same change of variables gives
\begin{align}
\int_0^\Lambda dE\,\rho(-E)P_\tau(-E)
&=
\frac{1}{\tau}
\int_0^{\tau\Lambda}
d\varepsilon\,
\rho(-\varepsilon/\tau)
P_\tau(-\varepsilon/\tau)
\nonumber\\
&\simeq
\frac{\rho_0}{\tau^{\nu+1}}
\int_0^{\tau\Lambda}
d\varepsilon\,
\varepsilon^\nu\mathcal P(-\varepsilon)
\nonumber\\
&\le
\frac{\rho_0}{\tau^{\nu+1}}
\int_0^\infty
d\varepsilon\,
\varepsilon^\nu\mathcal P(-\varepsilon).
\label{eqapp:gapless_dos_scaling}
\end{align}

When the remaining weighted contribution from $E\geq\Lambda$ is subleading, we obtain
\begin{equation}
\mathcal A_{\rm stop}(\tau;\mathcal S_{\rm gapless,\rho})
\sim \tau^{-(\nu+1)}.
\end{equation}
Physically, this scaling reflects the number of low-energy modes within the unresolved energy window $|E|\lesssim1/\tau$.

\begin{figure}[t]
\centering
\includegraphics[height=\paperfigureheight]{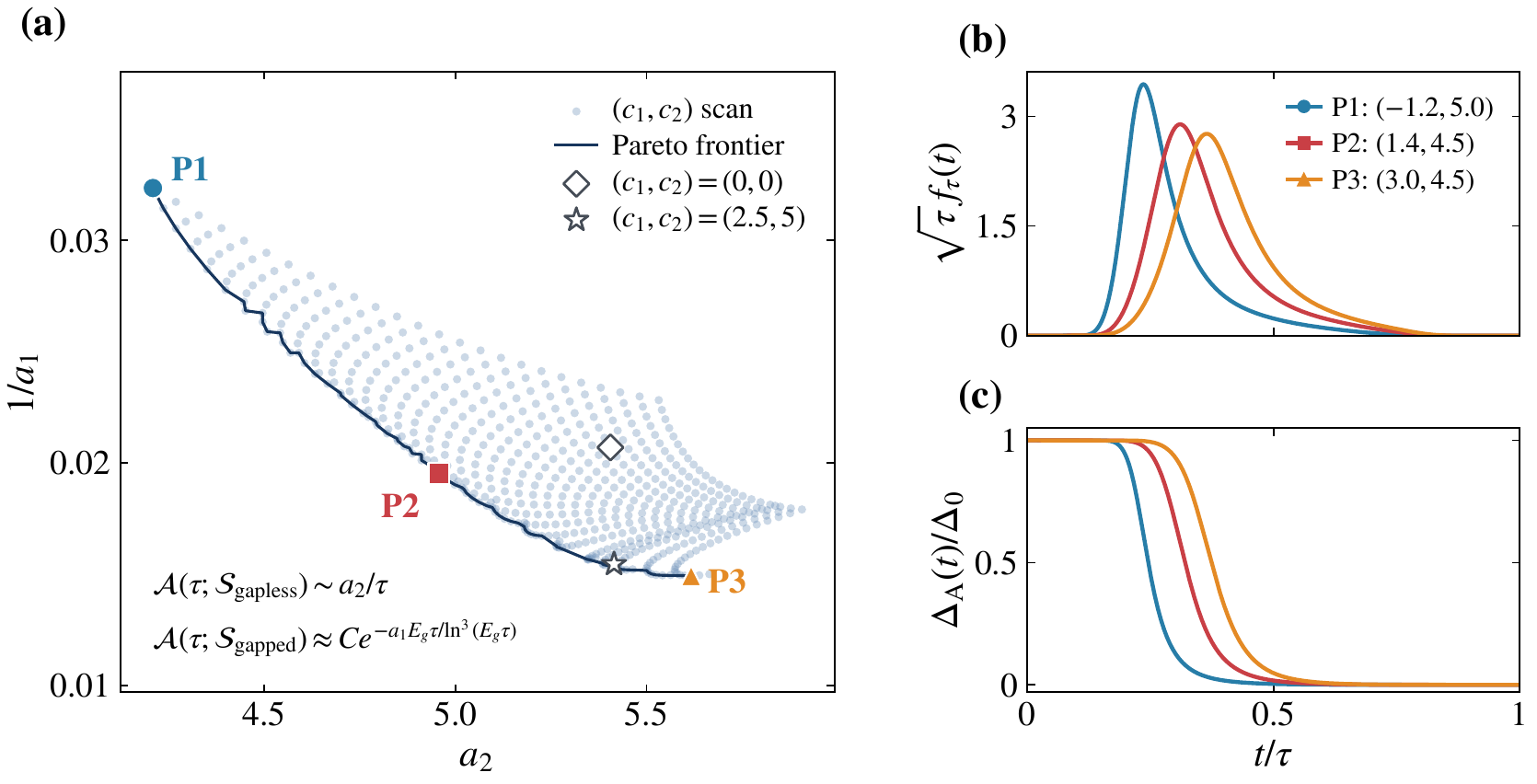}
\caption{
Coefficient trade-off within the M\"obius--Ingham family.
(a) Pareto frontier from different choices of $(c_1,c_2)$ with
$c_1\in[-1.2,3.0]$ and $c_2\in[-1,5]$.
The gapped coefficient $a_1$ is obtained from
$\log\mathcal A=\log C-a_1E_g\tau/\ln^3(E_g\tau)$,
with a free prefactor $C$.
The fits have $0.9843\leq R^2\leq0.9939$.
A pure-exponential fit over the same period range gives higher
$R^2$ values, all above $0.99$.
The gapless coefficient $a_2$ is the converged plateau
of $\tau\mathcal A(\tau;\mathcal S_{\mathrm{gapless}})$.
P1--P3 mark three representative frontier points, with parameters given in (b).
(b,c) Their normalized coupling functions $\sqrt{\tau}f_\tau(t)$ and
ancilla ramps $\Delta_{\mathrm A}(t)/\Delta_0$ at $\tau=60$ and
$\Delta_0=2.25$, satisfying the optimal matching condition.
Other parameters are as in
Fig.~\ref{fig:cooling_protocol}.}
\label{fig:coefficient_pareto}
\end{figure}

\section{M\"obius--Ingham control functions}
\label{subsec:reweighted_mobius_ingham_class}

We now give the construction introduced in Sec.~\ref{subsec:mobius_ingham_design}. Starting from the BMNS--Ingham function, we introduce a M\"obius transformation to further satisfy the optimal gapless scaling.

Let $s_{\mathrm I}(x)$ denote the BMNS--Ingham function defined in Appendix~\ref{App:inghamfunction}. We consider the two-parameter deformation
\begin{equation}
s_{\mathbf c}(x)
=
s_{\mathrm I}(x)e^{h_{\mathbf c}(x)},
\qquad
h_{\mathbf c}(x)
=
c_1(2x-1)+c_2(2x-1)^2,
\label{eq:reweighted_function_general}
\end{equation}
where $\mathbf c=(c_1,c_2)$. We further define the normalized integral of its square as
\begin{equation}
    Z_{\mathbf c}
    =
    \int_0^1 s_{\mathbf c}(u)^2\,du,
    \qquad
    A_{\mathbf c}(x)
    =
    \frac{1}{Z_{\mathbf c}}
    \int_0^x s_{\mathbf c}(u)^2\,du.
    \label{eq:reweighted_A}
\end{equation}
Applying the M\"obius transformation on $  A_{\mathbf c}(x)$ and using the optimal matching condition, one has

\begin{align}
\Delta_{\rm A;{\mathbf c}}(t)
&=
\Delta_0
\frac{1-A_{\mathbf c}(t/\tau)}
{1+(\Delta_0\tau-1)A_{\mathbf c}(t/\tau)},
\\
f_{\tau;{\mathbf c}}(t)
&=
\frac{\sqrt{\Delta_0}\,s_{\mathrm c}(t/\tau)}
{\sqrt Z_{\mathbf c}\,[1+(\Delta_0\tau-1)A_{\mathbf c}(t/\tau)]}.
\end{align}

\subsection{Gapped error scaling}
We first consider a gapped spectrum. The infinite endpoint flatness removes every finite-order boundary term in the first-order stopband error in Eq.~\eqref{eq:app_first_order_stopband}. The remaining error is determined by the Fourier tail of the function in Eq.~\eqref{eq:gapped_fourier_amplitude}. Because the M\"obius transformation depends on $\tau$, its derivatives must be estimated together with their period dependence.


\paragraph{Theorem 1: First-order almost-exponential stopband suppression.}
Consider the M\"obius--Ingham protocol at any fixed coupling strength $g_0>0$, $\mathbf c$, and a gapped spectrum $0<E_g\le E\le E_{\max}<\Delta_0$. The first-order stopband error
\begin{equation}
\mathcal A_{\rm stop}^{(1)}
(\tau;\mathcal S_{\rm gapped})
=
\int_{E_g}^{E_{\max}}
\left|c_{1,E}^{(1)}(\tau)\right|^2\,dE
\end{equation}
satisfies
\begin{equation}
\mathcal A_{\rm stop}^{(1)}
(\tau;\mathcal S_{\rm gapped})
\le
CW\exp\left[
-a_1\frac{E_g\tau}{\ln^3(E_g\tau)}
\right],
\label{eq:ingham_gapped_theorem}
\end{equation}
for $\tau\gg 1$. The constants $C,a_1>0$ are independent of $\tau$. See proof in Sec.~\ref{App:Proof of theorem 1}.

Neglecting higher-order nonadiabatic corrections, we approximate the exact stopband error by
\begin{equation}
\mathcal A_{\rm stop}
(\tau;\mathcal S_{\rm gapped})
\simeq
\mathcal A_{\rm stop}^{(1)}
(\tau;\mathcal S_{\rm gapped}).
\end{equation}
In the period range $We^{-\Gamma_0}\lesssim\mathcal A_{\rm stop}\ll W$, and under the passband correction condition in
Eq.~\eqref{eq:app_passband_correction_condition}, the total transfer error is of the same order as
the stopband error. Eq.~\eqref{eq:ingham_gapped_theorem} then gives
\begin{equation}
\mathcal A(\tau;\mathcal S_{\rm gapped})
=
O(\mathcal A_{\rm stop})
\lesssim
CW\exp\!\left[
-a_1\frac{E_g\tau}{\ln^3(E_g\tau)}
\right].
\label{eq:mobius_gapped_finite_period_bound}
\end{equation}
Comparing this bound with $We^{-\Gamma_0}$ gives an estimate of the upper period limit,
\begin{equation}
\tau^*
\sim
\frac{\Gamma_0}{a_1E_g}
\ln^3\!\left(\frac{\Gamma_0}{a_1}\right),
\qquad
\frac{\Gamma_0}{a_1}\gg1.
\end{equation}

\subsection{Gapless error scaling}

Notice that the denominators grow proportionally to $\tau$, so the rescaled control functions have the limits
\begin{align}
\tau\Delta_{\mathrm A;\mathbf c}(\tau x)
&\longrightarrow
\frac{1-A_{\mathbf c}(x)}{A_{\mathbf c}(x)},
\nonumber\\
\tau f_{\tau;\mathbf c}(\tau x)
&\longrightarrow
\frac{s_{\mathbf c}(x)}
{\sqrt{\Delta_0Z_{\mathbf c}}\,A_{\mathbf c}(x)}.
\label{eq:reweighted_gapless_limits}
\end{align}
 Hence, the corresponding Schr\"odinger equation becomes independent of $\tau$ at fixed $E\tau$, recovering the dominant gapless scaling discussed above 
\begin{align}
      \mathcal A(\tau;\mathcal S_{\rm gapless})
    \simeq
    \frac{a_2}{\tau},
 \end{align}
when the transition error dominates, i.e.,  $a_2/\tau\gg \max\{\mathcal A(\tau;\mathcal S_{\rm gapped}), W\exp(-\Gamma_0)\}$ with the almost-exponential decay gapped error discussed above.

\section{Single-passage cooling and Fourier-tail-induced additional crossings}
We give an intuitive argument for why suppressing the Fourier tail of the coupling function can improve cooling. In Sec.~\ref{sec:single_vs_multiple_LZ}, we show how multiple Landau--Zener passages within a single cooling cycle can reduce the cooling efficiency. In Sec.~\ref{Sec:Fouriertails}, we use a rotating-frame transformation to explain how nonzero-frequency components of the coupling can introduce additional resonances and transfer excitations back to the system, thereby increasing the cooling error.

\subsection{Single crossing versus multiple crossings in the Landau--Zener regime}
\label{sec:single_vs_multiple_LZ}

We now discuss why a single monotonic ramping is optimal in the strong-coupling Landau--Zener regime. For a monotonically decreasing ancilla energy, each mode $0<E<\Delta_0$ satisfies the resonance condition at $\Delta_{\mathrm A}(t_E)=E$ only once. Near this crossing, the remaining/transfer probability is
$|\alpha_-(E)|^2
\simeq
e^{-\Gamma_E}
$and $
|\beta_-(E)|^2
\simeq
1-e^{-\Gamma_E},$ respectively, where  $\Gamma_E
=
\frac{
2\pi g_0^2 V^2 |f(t_E)|^2
}{
\left|
\dot{\Delta}_{\mathrm A}(t_E)
\right|
}$ is introduced earlier.  Therefore, for a single crossing,
\begin{equation}
P_{\mathrm{cool}}^{(1)}(E)
=
|\beta_-(E)|^2.
\label{eq:single_crossing_cooling_probability}
\end{equation}

In the strong-coupling regime $\Gamma_E\gg1$, one has $|\alpha_-(E)|^2\rightarrow0 $ and $|\beta_-(E)|^2\rightarrow1,$
so that a single crossing produces an almost complete transfer of the excitation into the sink.

We next consider a non-monotonic protocol in which the same resonance is crossed twice, at
\begin{equation}
t_{E,1}<t_{E,2},
\qquad
\Delta_{\mathrm A}(t_{E,1})
=
\Delta_{\mathrm A}(t_{E,2})
=
E.
\end{equation}
There are two natural limits for this double-crossing process. In the first, coherence between the two passages is completely lost, so that the two crossings can be treated incoherently. In the second, coherence is fully preserved between the two passages, leading to Landau--Zener--Stueckelberg interference~\cite{Landau-Zener-Stinterferometry}.

We first consider the completely incoherent limit and classical analysis gives us the transfer probability
\begin{equation}
P_{\mathrm{cool,inc}}^{(2)}(E)
=
2P_{\mathrm{cool}}^{(1)}
\left[
1-P_{\mathrm{cool}}^{(1)}
\right].
\label{eq:double_crossing_incoherent}
\end{equation}
Comparing with the single-crossing result gives $P_{\mathrm{cool}}^{(1)}
-P_{\mathrm{cool,inc}}^{(2)} = P_{\mathrm{cool}}^{(1)} \left( 2P_{\mathrm{cool}}^{(1)}-1
\right).$
Hence,
\begin{equation}
P_{\mathrm{cool}}^{(1)}>\frac{1}{2}
\quad\Longrightarrow\quad
P_{\mathrm{cool}}^{(1)}(E)
>
P_{\mathrm{cool,inc}}^{(2)}(E).
\label{eq:single_better_incoherent}
\end{equation}

In particular, in the strong-coupling limit $P_{\mathrm{cool}}^{(1)}\rightarrow1,$ we have
\begin{equation}
P_{\mathrm{cool,inc}}^{(2)}(E)
\rightarrow0.
\label{eq:double_crossing_incoherent_strong}
\end{equation}
We can also consider the fully coherent limit, where the two transition paths interfere, giving rise to Landau--Zener--Stueckelberg interference. For two identical crossings, the standard double-passage Landau--Zener result gives ~ \cite{Landau-Zener-Stinterferometry}
\begin{equation}
P_{\mathrm{cool,coh}}^{(2)}(E)
=
4|\alpha_-(E)|^2|\beta_-(E)|^2
\sin^2\Phi_E,
\label{eq:double_crossing_coherent}
\end{equation}
where the Stueckelberg phase is $\Phi_E
=
\frac{1}{2}
\int_{t_{E,1}}^{t_{E,2}}
dt\,
\Omega_E(t)
+
\widetilde{\varphi}_{\mathrm S}(E)$ with $\Omega_E(t)
=
\sqrt{
\left[E-\Delta_{\mathrm A}(t)\right]^2
+
4|G(t)|^2
}.$ Here $\widetilde{\varphi}_{\mathrm S}$ is the shifted Landau--Zener Stokes phase.

Since $0\leq \sin^2\Phi_E\leq1,$ the coherent double-crossing probability obeys the phase-independent bound
\begin{equation}
P_{\mathrm{cool,coh}}^{(2)}(E)
\leq
4|\alpha_-(E)|^2|\beta_-(E)|^2.
\label{eq:double_crossing_coherent_bound}
\end{equation}
In the strong-coupling Landau--Zener limit,
\begin{equation}
|\alpha_-(E)|^2\rightarrow0,
\qquad
|\beta_-(E)|^2\rightarrow1,
\end{equation}
and therefore
\begin{equation}
P_{\mathrm{cool,coh}}^{(2)}(E)
\rightarrow0,
\qquad
P_{\mathrm{cool}}^{(1)}(E)
=
|\beta_-(E)|^2
\rightarrow1.
\label{eq:coherent_strong_coupling_limit}
\end{equation}
Thus, a second crossing becomes detrimental in the strong-coupling limit, independently of the Stueckelberg phase. The incoherent result in Eq.~\ref{eq:double_crossing_incoherent}  can be recovered by averaging over the Stueckelberg phase using $\left\langle
\sin^2\Phi_E
\right\rangle
=
\frac{1}{2}$.

\subsection{Additional crossings induced by Fourier tails in the gapped regime}~\label{Sec:Fouriertails}

At first glance, it may appear that our protocol naturally contains only a single crossing.  Indeed, the optimal matching condition in Eq.~\eqref{eq:general_optimal_velocity} requires $\Delta_{\mathrm A}(t)$ to decrease monotonically.  Therefore, for a fixed positive system energy $E$, the bare system and ancilla levels cross only once at $\Delta_{\mathrm A}(t_E)=E$.

However, this level-crossing picture does not account for the nonzero-frequency components of the coupling function $f_\tau(t)$.  A control function $f_\tau(t)$ of finite cycle period $\tau$ that is compactly supported in $t\in[0,\tau]$ cannot be strictly confined to zero frequency based on Paley–Wiener theorem and its Fourier transform necessarily contains nonzero-frequency components.

We now explain how these additional frequency components modify the resonance structure.  We focus on $H_E^-(t)$ in Eq.~\eqref{eq:app_HEminusappendix} and  the $H_E^+(t)$ sector can be analyzed similarly.  For simplicity, it is sufficient to consider a coupling with two representative frequency components,
\begin{equation}
f_\tau(t)
\simeq
f_0(t)
+
\epsilon f_1(t)e^{-i\omega t},
\qquad
|\epsilon|\ll1,
\label{eq:fourier_tail_toy_component}
\end{equation}
where $f_0(t)$ and $f_1(t)$ are real and vary slowly compared with the oscillatory factor $e^{-i\omega t}$.  The first term represents the dominant zero-frequency component, while the second term represents a nonzero-frequency component of the Fourier tail.

The zero-frequency component gives the usual resonance $E-\Delta_{\mathrm A}(t_E)=0.$ Now consider the nonzero-frequency component.  Its contribution to the coupling is
\begin{equation}
G_\omega(t)
=
\frac{\epsilon g_0 f_1(t)}{\sqrt{2}}
e^{-i\omega t}.
\end{equation}
Transforming to the frame rotating at frequency $\omega$ using $R_\omega(t) = e^{-i\omega t\sigma_z/2}$ gives
\begin{align}
H_{E,\omega}^{-,{\rm rot}}(t)
&=
R_\omega^\dagger
H_{E,\omega}^{-}(t)
R_\omega
-
i R_\omega^\dagger\dot R_\omega
\nonumber\\
&=
\begin{pmatrix}
\dfrac{E-\Delta_{\mathrm A}(t)- \omega}{2}
&
\dfrac{\epsilon g_0 f_1(t)}{\sqrt{2}}
\\ 
\dfrac{\epsilon^* g_0^* f_1(t)^*}{\sqrt{2}}
&
-\dfrac{E-\Delta_{\mathrm A}(t)- \omega}{2}
\end{pmatrix}.
\label{eq:fourier_tail_rotating_frame}
\end{align}
Thus, the nonzero-frequency component produces an additional resonance in the rotating frame when
\begin{equation}
E-\Delta_{\mathrm A}(t_\omega)
=\omega.
\label{eq:fourier_shifted_resonance}
\end{equation}
 Therefore, a system energy mode $E$ can effectively experience more than one crossing within a single cooling cycle due to the non-vanishing Fourier tail of $f_{\tau}(t)$.  Since these additional crossings are always detrimental to cooling as shown in Sec.~\ref{sec:single_vs_multiple_LZ}, we try to suppress the Fourier tail error as small as possible.

\section{Proof of Theorem 1}\label{App:Proof of theorem 1}

We first establish derivative bounds for the BMNS--Ingham function and show that the analytic reweighting and M\"obius transformation preserve the form of these derivative bounds. We then use these bounds and repeated integration by parts of the first-order Fourier representation to derive the almost-exponential stopband suppression.

\textit{1.Derivative bounds for the BMNS--Ingham function.}

We use the BMNS--Ingham function $s_{\mathrm I}(x)$ defined in Sec.~\ref{App:inghamfunction}.

Using $|t| \left|\frac{\sin(a_\ell t)}{a_\ell t} \right|^2 \le\frac{1}{a_\ell},$ we obtain
\begin{align}
\|s_{\mathrm I}^{(j)}\|_\infty
&\le
\frac{1}{2\pi}
\int_{\mathbb R}
|t|^j
\prod_{\ell=1}^{\infty}
\left(
\frac{\sin(a_\ell t)}{a_\ell t}
\right)^2dt
\nonumber\\
&\le
\frac{1}{2a_1}
\prod_{\ell=2}^{j+1}\frac{1}{a_\ell}
\nonumber\\
&\le
C^{j+1}j![\ln(e+j)]^{2j},j\ge 1
\label{eq:ingham_function_derivative_bound}
\end{align}

We next consider the analytic reweighting defined in Eq.~\eqref{eq:reweighted_function_general}.
Noting that the normalized function is $s(x)=
\frac{s_{\mathbf c}(x)}{\sqrt{Z_{\mathbf c}}}=\frac{s_{\mathrm I}(x)e^{h_{\mathbf c}(x)}}
{\sqrt{Z_{\mathbf c}}}$. Since $h_{\mathbf c}(x)$ is a quadratic polynomial, $e^{h_{\mathbf c}(x)}$ is entire. Cauchy's derivative estimate on a fixed complex neighborhood of $[0,1]$ gives
\begin{equation}
\left\|
\frac{d^j}{dx^j}e^{h_{\mathbf c}(x)}
\right\|_\infty
\le C^{j+1}j!,
\end{equation}
where $C$ depends only on the fixed reweighting parameters $\mathbf c$.

Leibniz's rule therefore gives
\begin{equation}
\|s^{(j)}\|_\infty
\le
C^{j+1}j![\ln(e+j)]^{2j}.
\label{eq:reweighted_function_derivative_bound}
\end{equation}

For the integral
\begin{equation}
A_{\mathbf c}(x)
=
\int_0^x s(u)^2\,du,
\qquad
A_{\mathbf c}'(x)=s(x)^2,
\end{equation}
Leibniz's rule gives, for $j\ge1$,
\begin{equation}
A_{\mathbf c}^{(j)}(x)
=
\sum_{r=0}^{j-1}
\binom{j-1}{r}
s^{(r)}(x)s^{(j-1-r)}(x).
\end{equation}
Using Eq.~\eqref{eq:reweighted_function_derivative_bound} and $\binom{j-1}{r}r!(j-1-r)!=(j-1)!$, we obtain
\begin{equation}
\|A_{\mathbf c}^{(j)}\|_\infty
\le
C^{j+1}j![\ln(e+j)]^{2j}.
\label{eq:A_derivative_bound}
\end{equation}
Thus, the analytic reweighting and integration preserve derivative bounds of the same form.

\textit{2.Derivative bounds for the M\"obius denominator.}

We denote $D_\tau(x)=1+(\mathcal T-1)A_{\mathbf c}(x)$, where $\mathcal T=\Delta_0\tau$. Since both the ramping and coupling functions involve $D_\tau^{-1}$, we need to control its higher derivatives, which can be expressed in terms of the relative derivatives $D_\tau^{(j)}/D_\tau$. To establish the corresponding bounds, we set
\begin{align}
    m=\lceil\ln(e+\mathcal T)\rceil,
\end{align}
and consider two ranges of derivative orders separately.

\textit{(1)} For $1\le j<m$, we use a local Taylor approximation to obtain a derivative bound proportional to $A_{\mathbf c}(x)+1/\mathcal T$. For each fixed $x\in[0,1]$, define
\begin{equation}
P_x(u)
=
\sum_{r=0}^{m-1}
\frac{A_{\mathbf c}^{(r)}(x)}{r!}
\left(-\frac{u}{m}\right)^r,
\qquad u\in[0,1].
\end{equation}
By Taylor's theorem and Eq.~\eqref{eq:A_derivative_bound},
\begin{align}
\left|
A_{\mathbf c}\left(x-\frac{u}{m}\right)
-P_x(u)
\right|
&\le
C\left[
\frac{C\ln^2(e+m)}{m}
\right]^m
\nonumber\\
&\le
\frac{1}{\mathcal T},
\label{eq:mobius_taylor_remainder}
\end{align}
for sufficiently large $\mathcal T$.

Since $A_{\mathbf c}$ is nonnegative and nondecreasing, we have $0\le A_{\mathbf c}(x-u/m)\le A_{\mathbf c}(x)$ for $u\in[0,1]$ and using
 Eq.~\eqref{eq:mobius_taylor_remainder}, the triangle inequality gives
\begin{align}
|P_x(u)|
&\le
\left|A_{\mathbf c}\left(x-\frac{u}{m}\right)\right|
+
\left|P_x(u)-A_{\mathbf c}\left(x-\frac{u}{m}\right)\right|
\nonumber\\
&\le
A_{\mathbf c}(x)+\frac{1}{\mathcal T}.
\end{align}
Therefore,
\begin{equation}
\max_{0\le u\le1}|P_x(u)|
\le
A_{\mathbf c}(x)+\frac{1}{\mathcal T}.
\end{equation}
Applying Markov's polynomial inequality and using $P_x^{(j)}(0)=(-1/m)^jA_{\mathbf c}^{(j)}(x)$, we obtain
\begin{align}
|A_{\mathbf c}^{(j)}(x)|
&=m^j|P_x^{(j)}(0)|
\nonumber\\
&\le
m^j(2m^2)^j
\max_{0\le u\le1}|P_x(u)|
\nonumber\\
&\le
(2m^3)^j
\left[
A_{\mathbf c}(x)+\frac{1}{\mathcal T}
\right].
\label{eq:mobius_fixed_scale_bound}
\end{align}
Consequently,
\begin{align}
\frac{|D_\tau^{(j)}(x)|}{j!D_\tau(x)}
&\le
(2m^3)^j
\frac{
(\mathcal T-1)
[A_{\mathbf c}(x)+1/\mathcal T]
}{
1+(\mathcal T-1)A_{\mathbf c}(x)
}
\nonumber\\
&\le
(2m^3)^j
\nonumber\\
&\le
[C\ln^3(e+\mathcal T)]^j.
\label{eq:mobius_relative_derivative_low}
\end{align}

\textit{(2)}For $m\le j\le\lfloor\mathcal T\rfloor$, the global derivative estimate is sufficient. Since $j\ge m\ge\ln\mathcal T$, we have $\mathcal T\le e^j$. Together with $j\le\mathcal T$ and $D_\tau\ge1$, this gives
\begin{align}
\frac{|D_\tau^{(j)}(x)|}{j!D_\tau(x)}
&\le
\mathcal T C^{j+1}
[\ln(e+j)]^{2j}
\nonumber\\
&\le
[C\ln^2(e+\mathcal T)]^j
\nonumber\\
&\le
[C\ln^3(e+\mathcal T)]^j.
\label{eq:mobius_relative_derivative_high}
\end{align}

Combining both cases yields
\begin{equation}
\frac{|D_\tau^{(j)}(x)|}{j!D_\tau(x)}
\le
[C\ln^3(e+\mathcal T)]^j,
\qquad
1\le j\le\lfloor\mathcal T\rfloor.
\label{eq:mobius_denominator_relative_bound}
\end{equation}

We now control the derivatives of $D_\tau^{-1}$. Differentiating $D_\tau D_\tau^{-1}=1$ and applying Leibniz's rule gives
\begin{equation}
(D_\tau^{-1})^{(j)}
=
-\frac{1}{D_\tau}
\sum_{r=1}^{j}
\binom{j}{r}
D_\tau^{(r)}
(D_\tau^{-1})^{(j-r)}.
\end{equation}
Using Eq.~\eqref{eq:mobius_denominator_relative_bound}, we prove by induction that
\begin{equation}
\frac{D_\tau(x)}{j!}
\left|(D_\tau^{-1})^{(j)}(x)\right|
\le
[2C\ln^3(e+\mathcal T)]^j.
\end{equation}
The case $j=0$ follows from $D_\tau D_\tau^{-1}=1$. Assuming the bound holds through order $n$, the Leibniz identity gives
\begin{align}
&\frac{D_\tau(x)}{(n+1)!}
\left|(D_\tau^{-1})^{(n+1)}(x)\right|
\nonumber\\
&\le
\sum_{r=1}^{n+1}
[C\ln^3(e+\mathcal T)]^r
[2C\ln^3(e+\mathcal T)]^{n+1-r}
\nonumber\\
&=
(2^{n+1}-1)
[C\ln^3(e+\mathcal T)]^{n+1}
\nonumber\\
&\le
[2C\ln^3(e+\mathcal T)]^{n+1}.
\end{align}
This completes the induction. Since $D_\tau(x)\ge1$,
\begin{equation}
\left\|(D_\tau^{-1})^{(j)}\right\|_\infty
\le
j![2C\ln^3(e+\mathcal T)]^j,
\qquad
0\le j\le\lfloor\mathcal T\rfloor.
\label{eq:mobius_inverse_derivative_bound}
\end{equation}

Finally, we use the inverse-denominator bound to control the physical functions $q_{\tau;\mathbf c}$ and $F_{\tau;\mathbf c}$ with
\begin{align}
q_{\tau;\mathbf c}(x)
&=
\frac{\Delta_{\mathrm A;\mathbf c}(\tau x)}{\Delta_0},
\label{eq:reweighted_mobius_q}
\\
F_{\tau;\mathbf c}(x)
&=
\sqrt{\tau}\,f_{\tau;\mathbf c}(\tau x).
\label{eq:reweighted_mobius_F}
\end{align}

Since
\begin{equation}
q_{\tau;\mathbf c}
=
\frac{\mathcal T D_\tau^{-1}-1}{\mathcal T-1},
\end{equation}
for $j\ge1$ and $\mathcal T\ge2$, we have
\begin{align}
\|q_{\tau;\mathbf c}^{(j)}\|_\infty
&=
\frac{\mathcal T}{\mathcal T-1}
\|(D_\tau^{-1})^{(j)}\|_\infty
\nonumber\\
&\le
Cj![C\ln^3(e+\mathcal T)]^j.
\end{align}
The case $j=0$ follows from $0\le q_{\tau;\mathbf c}\le1$.

Similarly, using $F_{\tau;\mathbf c}/\sqrt{\mathcal T} =sD_\tau^{-1}$ and Leibniz's rule, we obtain
\begin{align}
\left\|
\left(\frac{F_{\tau;\mathbf c}}{\sqrt{\mathcal T}}\right)^{(j)}
\right\|_\infty
&\le
\sum_{r=0}^j
\binom jr
\|s^{(r)}\|_\infty
\|(D_\tau^{-1})^{(j-r)}\|_\infty
\nonumber\\
&\le
Cj![C\ln^3(e+\mathcal T)]^j,
\label{eq:mobius_control_derivative_bound}
\end{align}
where we used $\binom jr r!(j-r)!=j!$ and $\ln(e+r)\le\ln(e+\mathcal T)$.

\textit{3. Derivative bounds for the Fourier transform.}

For the M\"obius--Ingham protocol, we define
\begin{equation}
\mathcal W_{E,\tau}(u)
=
\left.
\arctan\left[
\frac{
\sqrt{2\Delta_0}\,g_0s(x)
}{
D_\tau(x)
[E+\Delta_0q_{\tau;\mathbf c}(x)]
}
\right]
\right|_{x=X_{E,\tau}(u)}
\label{eq:mobius_fourier_profile}
\end{equation}
used in Eq.~\eqref{eq:gapped_weighted_oscillatory_integral}. Here $X_{E,\tau}(u)$ is the inverse of the exact phase variable defined in Eq.~\eqref{eq:gapped_plus_phase_variable}.

Since $du/dx=\Omega_E^+(\tau x)\ge E_g>0$, the change of variables is nonsingular. Applying finite-order composition and inversion estimates ~\cite{ultradifferentiablefunctionclass}, we obtain the following bound
\begin{equation}
\|\partial_u^n\mathcal W_{E,\tau}\|_{L^1}
\le
CE_g n!
\left[
\frac{C\ln^3(e+\mathcal T)}{E_g}
\right]^n,
\qquad
0\le n\le\lfloor\mathcal T\rfloor.
\label{eq:mobius_first_order_L1_bound}
\end{equation}

\textit{4. Fourier-tail suppression.}
Since $\mathcal W_{E,\tau}$ is infinitely flat at both endpoints, repeated integration by parts on Eq.~\eqref{eq:gapped_exact_oscillatory_integral} gives
\begin{align}
|c_{1,E}^{(1)}(\tau)|
&\le
CE_g\tau\,n!
\left[
\frac{C\ln^3(e+\mathcal T)}
{E_g\tau}
\right]^n,
\label{eq:mobius_first_order_ibp}
\end{align}
for $1\le n\le\lfloor\mathcal T\rfloor$.

Choosing $n=\left\lfloor\frac{E_g\tau}{4C\ln^3(e+\mathcal T)}\right\rfloor$  and using $n!\le n^n$, the factorial term is bounded by $4^{-n}$. For sufficiently large $\tau$, we have
\begin{equation}
|c_{1,E}^{(1)}(\tau)|^2
\le
C\exp\left[
-a_1\frac{E_g\tau}
{\ln^3(E_g\tau)}
\right],
\label{eq:mobius_first_order_probability_bound}
\end{equation}

Integrating over the gapped spectrum, we finally obtain
\begin{align}
\mathcal A_{\rm stop}^{(1)}
(\tau;\mathcal S_{\rm gapped})
&=
\int_{E_g}^{E_{\max}}
|c_{1,E}^{(1)}(\tau)|^2\,dE\le
CW\exp\left[
-a_1\frac{E_g\tau}
{\ln^3(E_g\tau)}
\right].
\end{align}

\section{Benchmark protocols and numerical methods}
\label{app:benchmark_protocols}

\begin{figure*}[t]
\centering
\includegraphics[height=\paperfigureheight]{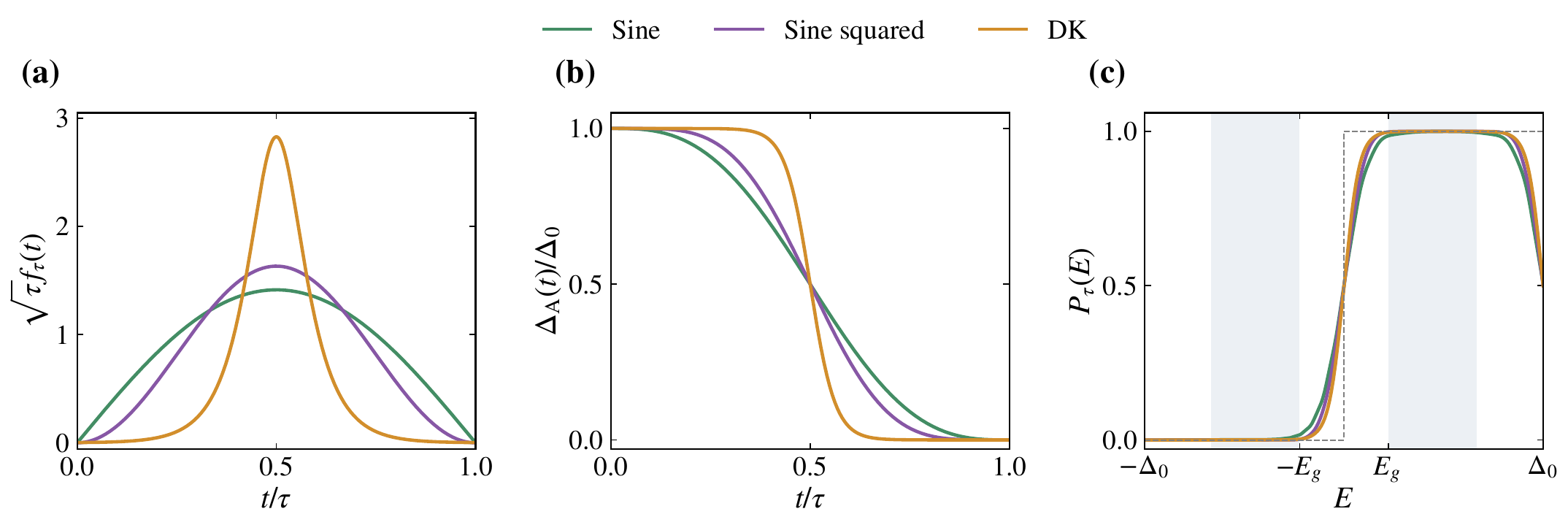}
\caption{Benchmark control functions at $\tau=60$. (a) Normalized coupling $\sqrt{\tau}f_\tau(t)$,
(b) ancilla ramp $\Delta_{\mathrm A}(t)/\Delta_0$,and (c) transfer probability $P_\tau(E)$ for the sine,sine-squared, and Demkov--Kunike (DK) protocols. All three satisfy the optimal matching condition in Eq.~\eqref{eq:optimal_matching_condition}. The DK protocol uses $r=16$. Other parameters are the same as Fig.~\ref{fig:cooling_protocol}.}
\label{fig:benchmark_protocols}
\end{figure*}

We summarize the benchmark control functions used in our comparisons. With $x=t/\tau\in[0,1]$, we have
\begin{equation}
f_\tau(t)=\frac{s(x)}{\sqrt{\tau}},
\qquad
\Delta_{\mathrm A}(t)=\Delta_0 q(x),
\end{equation}
where $\int_0^1s(x)^2\,dx=1$, $q(0)=1$, and $q(1)=0$. The optimal matching condition in Eq.~\eqref{eq:optimal_matching_condition} becomes $-q'(x)=s(x)^2$.

\subsection{BMNS--Ingham function}\label{App:inghamfunction}

Following the infinite-product construction of Bachmann et al.~\cite{Bachmanningham}, consider
\begin{equation}
w_\gamma(k)
=
c_\gamma
\prod_{\ell=1}^{\infty}
\left(
\frac{\sin(a_\ell k)}{a_\ell k}
\right)^2,
\qquad
a_\ell=\frac{a_1}{\ell(\ln\ell)^2}
\quad(\ell\ge2),
\label{eq:ingham_frequency_profile}
\end{equation}
where $\sum_{\ell\ge1}a_\ell=\gamma/2$. Taking $\gamma=1/2$, we define the shifted inverse Fourier transform
\begin{equation}
b_{\mathrm I}(x)
=
\frac{1}{2\pi c_\gamma}
\int_{\mathbb R}
w_\gamma(k)e^{ik(x-1/2)}\,dk.
\end{equation}
We define the normalized BMNS--Ingham function by
\begin{equation}
s_{\mathrm I}(x)
=
\frac{b_{\mathrm I}(x)}{\sqrt{Z_{\mathrm I}}},
\qquad
Z_{\mathrm I}
=
\int_0^1 b_{\mathrm I}(x)^2\,dx.
\label{eq:ingham_function_definition}
\end{equation}
The function $s_{\mathrm I}(x)$ is supported on $[0,1]$, symmetric about $x=1/2$, flat to all orders at both
endpoints, and satisfies $\int_0^1s_{\mathrm I}(x)^2\,dx=1$. The coupling function and ramp satisfying the optimal matching condition are
\begin{align}
f_\tau(t)
&=
\frac{1}{\sqrt{\tau}}s_{\mathrm I}(t/\tau),
\\
\Delta_{\mathrm A}(t)
&=
\Delta_0\left[
1-\int_0^{t/\tau}s_{\mathrm I}(u)^2\,du
\right],
\end{align}
for $0\leq t\leq\tau$. The functions are shown in Fig.~\ref{fig:cooling_protocol}(b,c).

\subsection{Sine and sine-squared control functions}
\label{app:sine_protocols}
For the sine control function, the normalized coupling and matched ramp are
\begin{equation}
s(x)=\sqrt{2}\sin(\pi x),\quad
q(x)=1-x+\frac{\sin(2\pi x)}{2\pi}.
\end{equation}
For the sine-squared control function,
\begin{align}
s(x)&=\sqrt{\frac{8}{3}}\sin^2(\pi x),\\
q(x)&=1-x+\frac{2\sin(2\pi x)}{3\pi}
-\frac{\sin(4\pi x)}{12\pi}.
\end{align}
Both satisfy $-q'(x)=s(x)^2$. Their first nonzero endpoint derivatives are of order $m=1$ and $m=2$, respectively, giving the first-order stopband scalings $\tau^{-3}$ and $\tau^{-5}$ discussed in
Appendix~\ref{app:gapped_switching} and shown in Fig.~\ref{fig:filter_error_scaling}(d).

\section{Gibbs-state preparation with the Demkov--Kunike protocol}\label{app:dk_thermal}

We first derive the approximate detailed balance condition for the Demkov--Kunike (DK) protocol defined on
$t\in(-\infty,+\infty)$ and then specify the finite-period DK protocol used in the numerics.

In our notation, Demkov--Kunike (DK) protocol in Ref.~\cite{demkov1969hypergeometric} is given by
\begin{align}
f_\infty(t)
&=
\frac{\operatorname{sech}(t/t_r)}{\sqrt{2t_r}},
\nonumber\\
\Delta_{\mathrm A,\infty}(t)
&=
\frac{\Delta_0}{2}[1-\tanh(t/t_r)],
\qquad t\in\mathbb R,
\label{eq:dk_full_pulse}
\end{align}
where $t_r$ is the width of the control functions. One can easily check that these controls satisfy the optimal matching condition.

Using the exact DK solution [Eq.~(41) of Ref.~\cite{ZlatanovVitanov2017}], the transition probability for the full protocol is
\begin{equation}
P_{\mathrm{DK}}(E)
=
\frac{\mathcal N_{\mathrm{DK}}}{
\cosh(\beta_{\mathrm{th}}\Delta_0/2)
+
\cosh[\beta_{\mathrm{th}}(E-\Delta_0/2)]
},
\label{eq:dk_exact_probability}
\end{equation}
where
\begin{align}
\beta_{\mathrm{th}}
&=\pi t_r,
\nonumber\\
\mathcal N_{\mathrm{DK}}
&=
\cosh(\beta_{\mathrm{th}}\Delta_0/2)
-
\cos\!\left[
\pi t_r
\sqrt{\frac{g_0^2}{t_r}-\frac{\Delta_0^2}{4}}
\right].
\label{eq:dk_probability_numerator}
\end{align}
Here $\mathcal N_{\mathrm{DK}}$ is positive and independent of $E$. It therefore cancels from the ratio of heating and cooling probabilities:
\begin{align}
\frac{P_{\mathrm{DK}}(-E)}{P_{\mathrm{DK}}(E)}
&=
\frac{
\cosh(\beta_{\mathrm{th}}\Delta_0/2)
+
\cosh[\beta_{\mathrm{th}}(E-\Delta_0/2)]
}{
\cosh(\beta_{\mathrm{th}}\Delta_0/2)
+
\cosh[\beta_{\mathrm{th}}(E+\Delta_0/2)]
}
\nonumber\\
&=
\frac{
\cosh[\beta_{\mathrm{th}}(\Delta_0-E)/2]
}{
\cosh[\beta_{\mathrm{th}}(\Delta_0+E)/2]
}
\nonumber\\
&=
e^{-\beta_{\mathrm{th}}E}
\frac{
1+e^{-\beta_{\mathrm{th}}(\Delta_0-E)}
}{
1+e^{-\beta_{\mathrm{th}}(\Delta_0+E)}
}.
\label{eq:dk_exact_balance_ratio}
\end{align}
For $0<E\leq E_{\max}<\Delta_0$, Eq.~\eqref{eq:dk_exact_balance_ratio} gives
\begin{equation}
\frac{P_{\mathrm{DK}}(-E)}{P_{\mathrm{DK}}(E)}
=
e^{-\beta_{\mathrm{th}}E}
\left[
1+O\!\left(
e^{-\beta_{\mathrm{th}}(\Delta_0-E_{\max})}
\right)
\right].
\label{eq:dk_approximate_balance}
\end{equation}
Thus, detailed balance holds up to exponentially small relative corrections when $\beta_{\mathrm{th}}(\Delta_0-E_{\max})\gg1$, without requiring a weak-coupling approximation.

Substituting Eq.~\eqref{eq:dk_exact_balance_ratio} into Eq.~\eqref{Eq:steadystaterate} gives
\begin{equation}
p_{E,\mathrm{DK}}^\infty
=
\frac{1}{e^{\beta_{\mathrm{th}}E}+1}
+
O\!\left(e^{-\beta_{\mathrm{th}}\Delta_0}\right).
\end{equation}
Thus, the stationary occupations approach the Gibbs values with exponentially small finite-bandwidth corrections.

In the numerical calculations, we define the width $t_r\equiv\tau/r$ and shift the control functions to center at $t=\tau/2$. Restricting  $t\in [0,\tau]$ and normalizing the control functions give
\begin{align}
f_\tau(t)
&=
\frac{
\operatorname{sech}[(t-\tau/2)/t_r]
}{
\sqrt{2t_r\tanh(r/2)}
},
\nonumber\\
\Delta_{\mathrm A}(t)
&=
\frac{\Delta_0}{2}
\left[
1-
\frac{
\tanh[(t-\tau/2)/t_r]
}{
\tanh(r/2)
}
\right].
\label{eq:dk_finite_period_controls}
\end{align}

The corresponding thermal target becomes
\begin{equation}
\beta_{\mathrm{th}}=\frac{\pi\tau}{r}.
\end{equation}

The coupling function, ancilla ramp, and corresponding transfer probability are shown in
Fig.~\ref{fig:benchmark_protocols}.

Finite-pulse truncation introduces boundary errors in addition to the exponentially small corrections of the full DK solution. From Eq.~\eqref{eq:dk_finite_period_controls}, the endpoint coupling satisfies
\begin{equation}
|f_\tau(0)|^2=|f_\tau(\tau)|^2
=\frac{r}{\tau\sinh r}
=\frac{\pi}{\beta_{\mathrm{th}}\sinh r}.
\label{eq:dk_endpoint_coupling}
\end{equation}
These nonzero endpoints produce algebraically decaying boundary errors. At fixed target $\beta_{\mathrm{th}}$, increasing $r$ lengthens the cycle while keeping $t_r=\beta_{\mathrm{th}}/\pi$ fixed. The pulse therefore becomes narrower relative to the cycle, with $t_r/\tau=1/r$, and the endpoint coupling is exponentially suppressed in $r$.

\begin{figure}[t]
\centering
\includegraphics[height=\paperfigureheight]{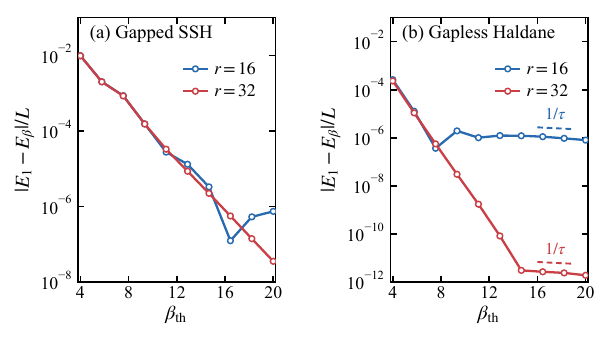}
\caption{Single-cycle Gibbs-energy error $|E_1-E_\beta|/L$ versus $\beta_{\mathrm{th}}$ for (a) a gapped SSH chain and (b) a gapless Haldane model, with $g_0\approx 4.2$ and $r=16,32$. Here $E_\beta$ is the target Gibbs energy. The error decreases exponentially with $\beta_{\mathrm{th}}$ until the endpoint errors become dominant.
Increasing $r$ makes the width of the coupling profile narrower relative to the cycle, suppressing these boundary errors and shifting the crossover to larger $\beta_{\mathrm{th}}$.
Dashed lines in (b) indicate $1/\tau$ reference scalings. Other parameters are the same as in Fig.~\ref{fig:Gibbs_state_preparation}.}
\label{fig:appendix_gibbs_energy_error}
\end{figure}

Fig.~\ref{fig:appendix_gibbs_energy_error} shows an initial  exponential decay with $\beta_{\mathrm{th}}$, consistent with the finite-bandwidth scale $e^{-\beta_{\mathrm{th}}(\Delta_0-E_{\max})}$ in Eq.~\eqref{eq:dk_approximate_balance}. Increasing $r$ suppresses endpoint errors and extends this exponential regime to larger $\beta_{\mathrm{th}}$.

\end{document}